\documentclass[prb,showpacs,draft,amsmath,amsfonts,twocolumn,superscriptaddress]{revtex4}

\usepackage[final]{graphicx}
\usepackage{bm}
\begin{document}

\title{Quantum many-body dynamics in a Lagrangian frame: II. Geometric
  formulation of time-dependent density functional theory.}

\author{I.~V.~Tokatly}
  
\email{ilya.tokatly@physik.uni-erlangen.de}

\affiliation{Lerhrstuhl f\"ur Theoretische Festk\"orperphysik,
  Universit\"at Erlangen-N\"urnberg, Staudtstrasse 7/B2, 91058
  Erlangen, Germany}

\affiliation{Moscow Institute of Electronic Technology,
 Zelenograd, 124498 Russia}
\date{\today}

\begin{abstract}
We formulate equations of time-dependent density functional theory
(TDDFT) in the co-moving Lagrangian reference frame. The main
advantage of the Lagrangian description of many-body dynamics is that
in the co-moving frame the current density vanishes, while the density
of particles becomes independent of time. Therefore a co-moving
observer will see the picture which is very similar to that seen in
the equilibrium system from the laboratory frame. It is shown that the
most natural set of basic variables in TDDFT includes the Lagrangian
coordinate, $\bm\xi$, a symmetric deformation tensor $g_{\mu\nu}$, and
a skew-symmetric vorticity tensor, $F_{\mu\nu}$.  These three
quantities, respectively, describe the translation, deformation, and the
rotation of an infinitesimal fluid element.  Reformulation of TDDFT in
terms of new basic variables resolves the problem of nonlocality and
thus allows to regularly derive a local nonadiabatic approximation for
exchange correlation (xc) potential. Stationarity of the density in
the co-moving frame makes the derivation to a large extent similar to
the derivation of the standard static local density approximation.  We
present a few explicit examples of nonlinear nonadiabatic xc
functionals in a form convenient for practical applications.
\end{abstract}


\maketitle

\section{Introduction}
Traditionally, the physical understanding of various many-body
phenomena is based on Landau's intuitive concept of
quasiparticles, which relates the behavior of a strongly interacting
quantum system to the properties of a gas of noninteracting (or weakly
interacting) elementary excitations \cite{PinNoz1}. The common
field-theoretical formulation of the many-body problem
\cite{AGD:e,FetWal,KadBay} allows to rigorously justify this very
appealing point of view, provided the system is in a weakly exited
many-body state. Unfortunately, practical applications of
traditional many-body methods to real systems, even in the equilibrium
or linear response regime, is computationally very demanding. In
strongly nonequilibrium systems the situation is much worse. The
simple intuitive picture of elementary excitations breaks down, while
the direct application of the standard many-body theory becomes
increasingly difficult even for model systems.

An alternative approach to the quantum many-body problem is offered by
the density functional theory (DFT)
\cite{DreizlerGross1990,BurkeGross1998,vanLeeuwen2001}. DFT represents
a mathematically rigorous realization of another famous idea in
theoretical physics which is a concept of collective variables
theory \cite{TerHaar}. Indeed, DFT opens a possibility to formulate
the many-body problem in a form of a closed theory that contains only
a restricted set of basic variables, such as density in the static
DFT, \cite{DreizlerGross1990} or density and current in the
time-dependent DFT (TDDFT)
\cite{RunGro1984,BurkeGross1998,vanLeeuwen2001}.  In classical physics
a theory of this type is known for more than 250 years. This is
classical hydrodynamics. In fact, the Runge-Gross mapping theorem in TDDFT
\cite{RunGro1984} proves the existence of an exact quantum
hydrodynamics. An analogy of TDDFT to hydrodynamics has been already
noted in the original paper by Runge and Gross,
Ref.~\onlinecite{RunGro1984} (see also our recent paper,
Ref.~\onlinecite{TokPPRB2003}). In this respect the static DFT should
be viewed as an exact quantum hydrostatics. It is indeed known that
the condition of energy minimum is equivalent to the condition for
a local compensation of the external and internal stress forces,
exerted on every infinitesimal element of an equilibrium system
\cite{BarPar1980}. Thus DFT not only introduces an alternative
formalism in the quantum many-body problem, but it also naturally suggests
an alternative way of thinking, which refers to the physical intuition,
developed over hundreds year of experience in the classical continuum
mechanics. Interestingly, equations of TDDFT in the hydrodynamic
formulation can be also considered as a force balance condition, but
in a local noninertial reference frame moving with the flow. In the
time-dependent case there is a local compensation of the external,
inertial, and internal stress forces. This demonstrates a close
similarity of the static DFT (which is currently a well developed
theory) and TDDFT (which is still under construction) in the co-moving
frame. The above similarity was the main motivation to reconsider the
formulation of TDDFT from the point of view of a local observer in the
co-moving Lagrangian reference frame. In this paper we present the
results of such a reconsideration, based on our recent formulation of
the many-body theory in the Lagrangian frame \cite{LagrFrameI} (in
what follows the papers of Ref.~\onlinecite{LagrFrameI} is referred to
as I).

Practical applications of any DFT rely on the Kohn-Sham (KS) construction
\cite{KohnSham1965,DreizlerGross1990}, which maps the calculation of
basic observables in the interacting system to the solution of an
auxiliary, noninteracting KS problem. Noninteracting KS particles move
in a self-consistent exchange correlation (xc) potential that is
adjusted to reproduce the correct values of basic variables,
i.~e. density and current in TDDFT. From the hydrodynamical point of
view the KS construction allows one to compute exactly the kinetic
part of the internal stress force, while treating the xc contribution
to the stress in an approximate fashion. Thus the central problem of any
practical DFT reduces to the construction of adequate approximations
for xc potentials. In the static DFT a good starting point is provided
by the local density approximation (LDA). On the one hand, the
static LDA, by itself, gives quite reasonable results, and, on the
other hand, it allows for further modifications and refinements. The
construction of a similar basic local approximation in TDDFT turned out to be
problematic. The reason for these problems is the inherent nonlocality
of the nonadiabatic nonequilibrium DFT.
General arguments \cite{Vignale1995a,Vignale1995b}, based on the
harmonic potential theorem \cite{Dobson1994}, require that any
consistent nonadiabatic xc potential must be a strongly nonlocal
functional of the density. Otherwise the harmonic potential theorem is
violated. 

In 1996 Vignale and Kohn \cite{VigKohn1996} (VK) realized that at
least in the linear response regime the problem of nonlocality can be
resolved by changing the basic variable and by using the xc vector
potential, ${\bf A}^{\text{xc}}$, instead of the scalar one. Namely, VK
showed that if one considers the current ${\bf j}$ (instead of the
density $n$) as a basic variable, a consistent linear local
approximation for ${\bf A}^{\text{xc}}$ can be regularly derived
\cite{VigKohn1996}. Shortly after that, Vignale, Ullrich and Conti
\cite{VigUllCon1997} (VUC) found that the further switching of
variables from the current ${\bf j}$ to the velocity ${\bf v}={\bf j}/n$,
allows to represent the complicated VK expression in a physically
transparent viscoelastic form. VUC also phenomenologically constructed
a nonlinear nonadiabatic xc functional. In this construction they
formally adopted the linear VK form, but with coefficients, taken at
a ``retarded position'' \cite{DobBunGro1997}. A similar, linear in
velocity gradients, phenomenological construction based on
Keldysh-contour action functional, has been recently proposed by
Kurzweil and Baer \cite{KurBae2004}. An attempt to regularly
derive a nonlinear nonadiabatic local approximation in TDDFT was made
in our recent work \cite{TokPPRB2003}. Noting that the applicability
conditions for the linear VK approximation exactly coincide with those
for the ``collisionless hydrodynamics'' of
Refs.~\onlinecite{TokPPRB1999,TokPPRB2000,AtwAshc2002}, we extended
the hydrodynamics formalism to TDDFT. Using Landau Fermi-liquid theory
we derived a nonadiabatic nonlinear approximation for the xc stress
tensor that defines the xc potential \cite{TokPPRB2003}. It has been
shown that the stress tensor is a local functional of new basic
variables: the Lagrangian coordinate and a second-rank metric-like
tensor. An appearance of the Lagrangian coordinate as a basic variable
is quite remarkable if we recall the abovementioned (see also I)
static-like force balance in the co-moving Lagrangian frame.

In this paper we reformulate TDDFT using the exact equations of
quantum many-body dynamics in the Lagrangian frame
\cite{LagrFrameI}. We show that possibly the most natural complete set
of basic variables in TDDFT consists of the Lagrangian coordinate
$\bm\xi$, the symmetric Green's deformation tensor $g_{\mu\nu}$, and
the skew-symmetric vorticity tensor $\widetilde{F}_{\mu\nu}$. These
three quantities, one vector, one symmetric and one skew-symmetric
tensors contain twelve numbers that are required for the complete
local characterization of a deformed state of any continuum media
\cite{PhysAc}. Namely, $\bm\xi$, $g_{\mu\nu}$ and
$\widetilde{F}_{\mu\nu}$ respectively describe the translation,
deformation and the rotation of an infinitesimal fluid element. On the
other hand, tensors $g_{\mu\nu}$ and $\widetilde{F}_{\mu\nu}$ describe
generalized inertia forces in the Lagrangian frame (see I). All three
quantities are functionals of velocity in accordance with the Runge-Gross
mapping theorem \cite{RunGro1984}. The new formulation of TDDFT
relates the local stress in the system to the dynamic deformations,
which is very natural physically. The main practical advantage is that
the reformulation of TDDFT in terms of new basic variables resolves
the problem of nonlocality on the most fundamental level. We show that
the dynamic xc~stress--deformation relation can be made local. This
allows us to derive a local nonadiabatic approximation in a regular
way that is similar to the derivation of the static LDA. The whole
history of nonadiabatic approximations in TDDFT can be viewed as a
staircase of successive transformations of basic variables, from the
density, via the current and the velocity to the general geometric
characteristics of deformed media. The first step was made by VK
\cite{VigKohn1996} in 1996. Hopefully the present formulation of TDDFT
corresponds to the last step on this staircase.

The structure of the paper is the following. In Sec.~II we consider the
hydrodynamic formulation of TDDFT. Using this formulation we introduce
the KS system and define xc potentials in terms of stress tensors. In
Sec.~III we discuss local approximations in DFT. We derive the common
static LDA and explain physical reasons of the nonlocality in TDDFT.
Sections IV and V are the central parts of the present paper. In
Sec.~IV we introduce new basic variables and develop a complete
geometric formulation of TDDFT. We also formulate a universal
many-body problem which determines the xc stress tensor and the xc
potential as functionals of basic variables.  In Sec.~V we discuss an
approximate solution of the universal many-body problem in the lowest
order of the gradient expansion.  Explicit examples of nonadiabatic
local xc functionals are presented in Secs.~VB--VD. In the concluding
part of Sec.~VD we formulate the full set of time-dependent KS
equations in the nonlinear time-dependent local deformation
approximation. In equilibrium case these equations reduce to the
common equations of DFT with the static LDA xc potential, while in the
linear response regime we recover the results of VK approximation.  In
Sec.~VI we summarize our results. Three appendixes contain technical
details of calculations.

\section{Definition of exchange correlation potentials via stress density}

In this section we discuss a hydrodynamic formulation of DFT and
introduce a definition of xc potentials in terms of local stress
forces. Let us consider a system of N interacting fermions in the
presence of a time-dependent external potential 
$U_{\text{ext}}({\bf x},t)$. The system is described by the following
Hamiltonian
\begin{equation}
H = \widehat{T}+\widehat{W}+\widehat{U},
\label{1}
\end{equation}
where $\widehat{T}$, $\widehat{W}$, and $\widehat{U}$ correspond to the
operators of the kinetic energy, interparticle interaction energy, and
the energy of interaction with the external field respectively
\begin{eqnarray}
\widehat{T}&=&
-\int d{\bf x}\psi^{\dag}({\bf x})\frac{\nabla^{2}}{2m}\psi({\bf x}),
\label{2}\\
\widehat{W} &=& \frac{1}{2}\int d{\bf x}d{\bf x'}w(|{\bf x-x'}|)
\psi^{\dag}({\bf x})\psi^{\dag}({\bf x'})\psi({\bf x'})\psi({\bf x}),
\label{3}\\
\widehat{U}&=&
\int d{\bf x}U_{\text{ext}}({\bf x},t)\psi^{\dag}({\bf x})\psi({\bf x}).
\label{4}
\end{eqnarray}
Here  $w(|{\bf x}|)$ is the potential of a pairwise interparticle
interaction.  Field operators $\psi^{\dag}$
and $\psi$ satisfy the fermionic commutation relation
\begin{equation}
\{\psi^{\dag}({\bf x}),\psi({\bf x'})\}=\delta({\bf x-x'}).
\label{5}
\end{equation}
For definiteness we consider a Fermi system, although most of the
results of this paper are independent of statistics.

The hydrodynamic formulation of DFT follows the Heisenberg equations of
motion for the density and for the current operators
\begin{eqnarray}
\frac{\partial}{\partial t}n({\bf x},t) 
- i\langle[H,\widehat{n}({\bf x})]\rangle &=& 0,
\label{6}\\
\frac{\partial}{\partial t}{\bf j}({\bf x},t)
- i\langle[H,\widehat{\bf j}({\bf x})]\rangle &=& 0,
\label{7}
\end{eqnarray}
where $n({\bf x})=\langle\widehat{n}({\bf x})\rangle$ and 
${\bf j}({\bf x})=\langle\widehat{\bf j}({\bf x})\rangle$ are the
density of particles and the current density. The corresponding
operators are defined by the standard expressions
\begin{eqnarray}
\widehat{n}({\bf x},t) &=& \psi^{\dag}({\bf x},t)\psi({\bf x},t),
\label{8}\\
\widehat{j}_{\mu}({\bf x},t) &=& -\frac{i}{2m}\left(
\psi^{\dag}\frac{\partial\psi}{\partial x^{\mu}} -
\frac{\partial\psi^{\dag}}{\partial x^{\mu}}\psi\right).
\label{9}
\end{eqnarray}
Angle brackets in the above formulas denote averaging with the exact
density matrix $\widehat{\rho}$, i.e. 
$\langle A\rangle = \text{Tr}\widehat{\rho}A$. Equations (\ref{6}),
(\ref{7}) can be represented in a form of the hydrodynamics
balance equations (details of the derivation can be found, for
example, in I)
\begin{eqnarray}
&&D_{t}n + n\frac{\partial}{\partial x^{\mu}}v_{\mu} = 0,
\label{10}\\
&&mnD_{t}v_{\mu} + \frac{\partial}{\partial x^{\nu}}P_{\mu\nu} +
n\frac{\partial}{\partial x^{\mu}}U=0,
\label{11}
\end{eqnarray}
Equation (\ref{10}) is the common continuity equation, while Eq.~(\ref{11})
corresponds to the local momentum conservation law. In these equations
${\bf v}={\bf j}/n$ is the velocity of the flow,  
$D_{t}=\frac{\partial}{\partial t} + {\bf v}\nabla$ is the
convective derivative and $U=U_{\text{ext}}+U_{\text{H}}$ is the sum
of the external and the Hartree potentials, 
\begin{equation}
U_{\text{H}}({\bf x},t) = \int w(|{\bf x-x'}|)n({\bf x'},t)d{\bf x'}.
\label{12}
\end{equation}
The exact stress tensor in Eq.~(\ref{11}),  
\begin{equation}
P_{\mu\nu}({\bf x},t) = T_{\mu\nu}({\bf x},t)+W_{\mu\nu}({\bf x},t),
\label{13a}
\end{equation}
contains the kinetic, $T_{\mu\nu}$, and the
interaction, $W_{\mu\nu}$, contributions. Divergences of the tensors
$T_{\mu\nu}$ and $W_{\mu\nu}$ in Eq.~(\ref{11}) come from the
commutators of the current operator with operators $\widehat{T}$,
Eq.~(\ref{2}), and $\widehat{W}$, Eq.~(\ref{3}), respectively. In I we
have derived the following explicit representations for the stress
tensors (see also
Refs.~\onlinecite{MartSchw1959,KadBay,PufGil1968,Zubarev:e}) 
\begin{eqnarray}
T_{\mu\nu}({\bf x})&=&\frac{1}{2m}\langle
(\widehat{q}_{\mu}\psi)^{\dag}\widehat{q}_{\nu}\psi +
(\widehat{q}_{\nu}\psi)^{\dag}\widehat{q}_{\mu}\psi -
\frac{\delta_{\mu\nu}}{2}\nabla^{2}\widehat{n}
\rangle\qquad
\label{13}\\
W_{\mu\nu}({\bf x}) &=& -\frac{1}{2}\int d{\bf x'}
\frac{{x'}^{\mu}{x'}^{\nu}}{|{\bf x'}|}
\frac{\partial w(|{\bf x'}|)}{\partial |{\bf x'}|}
\nonumber \\
&&\times \int_{0}^{1}
G_{2}({\bf x}+\lambda{\bf x'},{\bf x}-(1-\lambda){\bf x'})
d\lambda
\label{14}
\end{eqnarray}
where $\widehat{\bf q}=-i\nabla - m{\bf v}$ is the operator of
``relative'' momentum, and $G_{2}({\bf x},{\bf x'}) = 
\langle\psi^{\dag}({\bf x})\widehat{n}({\bf x'})\psi({\bf x})\rangle 
- n({\bf x})n({\bf x'})$ is the pair correlation function
\cite{note0}. It is worth mentioning that the representability of the
stress force in a form of a divergence of a tensor is a direct
consequence of the Newton's third law (see Appendix~A in I).

Equations (\ref{10}) and (\ref{11}) represent the exact local
conservation laws which must be satisfied for an arbitrary evolution
of the system. Let us apply them to TDDFT. The first, less restrictive
part of TDDFT mapping theorem \cite{RunGro1984,LiTon1985} states the
existence of a unique and invertible map: ${\bf j}\to U_{\text{ext}}$
or, equivalently, ${\bf v}\to U_{\text{ext}}$. This implies that the
exact many-body density matrix $\widehat{\rho}(t)$ for a given initial
condition, $\widehat{\rho}(0)=\widehat{\rho}_{0}$, is a functional of
the velocity ${\bf v}$. Hence the stress tensor of Eqs.~(\ref{13a}) is
a functional of ${\bf v}$ and of the initial density matrix:
$P_{\mu\nu}[\widehat{\rho}_{0},{\bf v}]$. Therefore Eqs.~(\ref{10})
and (\ref{11}) constitute a formally closed set of the exact quantum
hydrodynamics equations with the memory of initial many-body
correlations. It is interesting to note that the common 
classical hydrodynamics can be viewed as a particular limiting form of
TDDFT. In this limiting case the stress tensor functional is known
explicitly -- it takes the usual Navier-Stokes form \cite{LandauVI:e}.

In the equilibrium system Eq.~(\ref{11}) reduces to the static force
balance equation
\begin{equation}
\frac{\partial}{\partial x^{\nu}}P_{\mu\nu} +
n\frac{\partial}{\partial x^{\mu}}U = 0
\label{15}
\end{equation}
This equation shows that the force produced by the external and the Hartree
potentials is compensated by the force of internal stresses. The net
force, exerted on every infinitesimal fluid element is zero, which
results in zero current density and a stationary particles' density
distribution. According to the Hohenberg-Kohn theorem
\cite{HohKohn1964,DreizlerGross1990} any equilibrium observable,
in particular the stress tensor, is a functional of the density
$n$: $P_{\mu\nu}=P_{\mu\nu}[n]$. Hence Eq.~(\ref{15}) is, in fact, the
equation of the exact 
quantum hydrostatics that uniquely determines the density
distribution in the equilibrium system. In the semiclassical limit
Eq.~(\ref{15}) reduces to the common hydrostatics equation
\cite{LandauVI:e} (for a degenerate high density Fermi
gas we recover the Thomas-Fermi theory).

Practical application of DFT always rely on the KS construction that is
a particular consequence of the mapping theorems. The current and the
density in the interacting system can be reproduced in a system of
noninteracting KS particles, moving in a properly chosen self-consistent
potential $U_{\text{S}}=U + U_{\text{xc}}$ \cite{note1}. The hydrodynamic
formulation of TDDFT/DFT allows us to relate the xc potential
$U_{\text{xc}}$ to the stress density. Hydrodynamics balance equations
for the KS system take the form
\begin{eqnarray}
&&D_{t}n + n\frac{\partial}{\partial x^{\mu}}v_{\mu} = 0,
\label{16}\\
&&mnD_{t}v_{\mu} + \frac{\partial}{\partial x^{\nu}}T^{\text{S}}_{\mu\nu} +
n\frac{\partial}{\partial x^{\mu}}U_{\text{xc}} +
n\frac{\partial}{\partial x^{\mu}}U=0
\label{17}
\end{eqnarray}
where the kinetic stress tensor of KS system, $T^{\text{S}}_{\mu\nu}$,
is given by Eq.~(\ref{13}), but with the averaging over the state of
noninteracting particles. Comparing Eq.~(\ref{10}) and (\ref{11}) with
Eqs.~(\ref{16}) and (\ref{17}) we find that the velocity ${\bf v}$ and
the density $n$ of the noninteracting and the interacting systems
coincide if the xc potential $ U_{\text{xc}}({\bf x},t)$ satisfies 
the equation
\begin{equation} 
\frac{\partial U_{\text{xc}}}{\partial x^{\mu}} = 
\frac{1}{n}\frac{\partial P^{\text{xc}}_{\mu\nu}}{\partial x^{\nu}}
\label{18}
\end{equation}
where $P^{\text{xc}}_{\mu\nu}$ is the xc stress tensor that equals to
the difference of the stress tensors in the 
interacting and noninteracting systems with the same velocity
distribution
\begin{equation} 
P^{\text{xc}}_{\mu\nu} = P_{\mu\nu} - T^{\text{S}}_{\mu\nu}
\label{19}
\end{equation}
Equations (\ref{18}) and (\ref{19}) demonstrate the physical
significance of $U_{\text{xc}}$. The xc potential should produce a
force which compensates the difference of the internal stress forces in the
real interacting system and in the auxiliary noninteracting KS system.
By continuity equation, the density $n$ is a functional
of the velocity. Therefore Eq.~(\ref{18}) defines
$U_{\text{xc}}$ (up to inessential constant) as a
functional of ${\bf v}$. Equation (\ref{18}) shows that 
$n^{-1}\partial_{\nu}P^{\text{xc}}_{\mu\nu}$ is a potential
vector. This does not mean that vectors 
$n^{-1}\partial_{\nu}P_{\mu\nu}$ and 
$n^{-1}\partial_{\nu}T^{\text{S}}_{\mu\nu}$ separately have no
rotational components. However, according to the balance equations of
Eqs.~(\ref{11}) and (\ref{17}) the rotational components of these
vectors coincide -- both equal to the rotational part of the vector 
$mD_{t}{\bf v}$. 

It is also possible to
construct the proper KS system using the xc vector potential ${\bf
A}^{\text{xc}}$ or, in the
most general case, a combination of xc vector and scalar potentials
\cite{VigKohn1996,VigUllCon1997,TokPPRB2003,VigRas1988}. In this 
case the exact local conservation laws require that the total xc
force, ${\bm{\mathcal{F}}}^{\text{xc}}$, should compensate the
difference of the stress forces in the interacting and the KS systems:
\begin{equation}
{\cal F}^{\text{xc}}_{\mu} = 
\frac{\partial A^{\text{xc}}_{\mu}}{\partial t}
- ({\bf v}\times(\nabla\times{\bf A}^{\text{xc}}))_{\mu} 
+ \frac{\partial U_{\text{xc}}}{\partial x^{\mu}} = 
\frac{1}{n}\frac{\partial P^{\text{xc}}_{\mu\nu}}{\partial x^{\nu}}.
\label{18a}
\end{equation}
This equation determines the xc potentials, 
${\bf A}^{\text{xc}}$ and $U_{\text{xc}}$, up to a gauge
transform. Equation (\ref{18a}) represents a very important 
{\em exact} property of xc potentials: they produce a force that must be a
divergence of a second rank tensor. This requirement automatically
implies the well known zero force and zero torque sum rules
\cite{BurkeGross1998}    
\begin{equation}
\int n\bm{\mathcal{F}}^{\text{xc}}d{\bf x}=0, \quad
\int n({\bf x}\times\bm{\mathcal{F}}^{\text{xc}})d{\bf x}=0.
\label{18b}
\end{equation} 
We would like to outline that the exact local condition of
Eq.~(\ref{18a}) is much stronger then the common integral requirements
of Eq.~(\ref{18b}). Apparently the above definition of xc potentials
equally well apply both to TDDFT and to the static/equilibrium DFT. It
should be mentioned that in the equilibrium case the stress forces in
the interacting and KS systems separately are potential vectors. 

Let us briefly discuss the role of xc vector potential in
DFT. Apparently an appearance of ${\bf A}^{\text{xc}}$ is unavoidable
in the presence of an external magnetic field
\cite{VigRas1988}. Independently of the character of external fields,
the formulation in terms of ${\bf A}^{\text{xc}}$ is convenient in the
linear response regime \cite{VigKohn1996,VigUllCon1997}. Indeed, in
the linearized theory we can perform the Fourier transform in the time
domain, which makes ${\bf A}_{\text{xc}}$ completely local, provided
the xc stress tensor is a local functional of some basic
variables. In the nonlinear regime this advantage clearly
disappears. For a nonlinear evolution the description of xc
effects in terms of the scalar potential, defined by Eqs.~(\ref{20}),
(\ref{21}), looks at least as convenient as the formulation in terms
of ${\bf A}^{\text{xc}}$. Below for definiteness
we assume the noninteracting $v$-representability of the velocity, which
allows us to construct the KS system using only the scalar xc
potential. Reformulation of the theory for xc vector potential is
straightforward (see Conclusion).

For the practical applications it is possibly more convenient to
represent the force definition of $U_{\text{xc}}$, Eq.~(\ref{18}), in
a familiar form of the Poisson equation
\begin{equation} 
\nabla^{2}U_{\text{xc}}({\bf x},t) = 4\pi\rho_{\text{xc}}({\bf x},t),
\label{20}
\end{equation}
where the quantity $\rho_{\text{xc}}({\bf x},t)$,
\begin{equation} 
\rho_{\text{xc}} = \frac{1}{4\pi}\frac{\partial}{\partial x^{\mu}}
\left( \frac{1}{n}
\frac{\partial}{\partial x^{\nu}} P^{\text{xc}}_{\mu\nu}\right),
\label{21}
\end{equation}
can be interpreted as an xc ``charge'' density. In this context the xc
stress force, $n^{-1}\partial_{\nu}P^{\text{xc}}_{\mu\nu}$, has a
clear meaning of an xc ``polarization'' density. The additional
differentiation in Eq.~(\ref{20}) requires an additional boundary
condition. The most natural condition, which we should impose on the
solution to Eq.~(\ref{20}), is the requirement of boundness at
infinity.

Equation (\ref{18}) or, equivalently, Eqs.~(\ref{20}), (\ref{21})
reduce the problem of approximations for $U_{\text{xc}}$ to the
construction of approximations for the xc stress tensor
$P^{\text{xc}}_{\mu\nu}$. Since the  stress density has a clear
physical and microscopic meaning there is a hope that the later problem
is more tractable.

\section{Static LDA vs. time-dependent LDA}

Let us first derive the standard static LDA from
the force definition of $U_{\text{xc}}$, Eq.~(\ref{18}). Formally the
static $U_{\text{xc}}^{\text{LDA}}({\bf x})$ is the solution to
Eq.~(\ref{18}) in the 
lowest order of the gradient expansion. This solution is obtained by
inserting
$P^{\text{xc}}_{\mu\nu}$ for a homogeneous system of the density
$n({\bf x})$ into the right hand side of Eq.~(\ref{18}). In the
homogeneous system the stress 
tensors $P_{\mu\nu}$ and $T^{\text{S}}_{\mu\nu}$ are diagonal
$$  
P_{\mu\nu}[n] = \delta_{\mu\nu}P(n), \qquad 
T^{\text{S}}_{\mu\nu}[n] = \delta_{\mu\nu}P_{0}(n),
$$
where $P$ and $P_{0}$ are the pressure of the interacting system and
the pressure of an ideal gas respectively. Therefore to the lowest
order in the density gradients we get
\begin{equation}
P^{\text{xc}}_{\mu\nu}[n]({\bf x}) = 
\delta_{\mu\nu}P_{\text{xc}}(n({\bf x})),
\label{22}
\end{equation}
where $P_{\text{xc}}=P-P_{0}$ is the xc pressure of the homogeneous
system. Substituting Eq.~(\ref{22}) into Eq.~(\ref{18}), and using the
common thermodynamic relations, $dP = nd\mu$, 
$\mu = \partial F/\partial n$, we find the following result for the xc
potential
\begin{equation}
U_{\text{xc}}[n]({\bf x})= U_{\text{xc}}^{\text{LDA}}(n({\bf x})),
\qquad
U_{\text{xc}}^{\text{LDA}}(n) = 
\frac{\partial F_{\text{xc}}}{\partial n}.
\label{23}
\end{equation}
Here $F_{\text{xc}}$ is the xc free energy of the homogeneous
system. The result of Eq.~(\ref{23}) recovers the standard static LDA
\cite{DreizlerGross1990}. 

Physically the above derivation of the static LDA means the following. If
the density distribution is a semiclassically slow function in space,
every small volume element can be formally considered as an independent
homogeneous many-body system. The density in this homogeneous system
equals to the density at the location of the element. By solving the
homogeneous many-body problem we find the stress tensor, which, after
the substitution into Eq.~(\ref{18}) provides us with
$U_{\text{xc}}^{\text{LDA}}$.

The situation in the time-dependent theory is much more
complicated. Even if at any instant $t$ the density distribution
$n({\bf x},t)$ is a slow function in space, a small volume element, located at
some point ${\bf x}$, can not be considered as system that is
independent of surrounding space. For a nonadiabatic dynamics,
particles, arriving at the point ${\bf x}$ from other regions, bring an
information about other spatial points. This is the physical reason
for the well known nonlocality, immanent to any nonadiabatic TDDFT
\cite{Vignale1995a,Vignale1995b}. It is straightforward to demonstrate
the failure of any plain attempt to extend the above derivation of the
static LDA to the time-dependent case. Indeed, a homogeneous many-body
problem, which we would get by formally separating a small volume
element, corresponds to an infinite system with strongly nonconserved number of
particles. Apparently this problem is meaningless.

In the rest of this paper we show that the nonlocality problem in
TDDFT is resolved by changing a ``point of view'' on the nonequilibrium
many-body system. Any flow in the system can be considered as a
collection of small fluid elements moving
along their own trajectories. It is possible to divide the system
into elements in such a way that the number of particles in every element
will be conserved. Indeed, by the proper deformation and
rotation of a fluid element one can always adjust its shape to the
motion of particles and thus prevent the flow
through its surface. Let us attach a reference frame to one of
those moving elements. The motion of the origin of this frame compensates the
translational motion of the fluid element. By properly changing scales and
directions of coordinate axes we can also compensate both the
deformations and the rotation. This means that an observer in the new
frame will see no currents in the system, and a stationary density
distribution. Thus from the point of view of the co-moving observer the
nonequilibrium system looks very similar to the equilibrium one, as it
is seen by a stationary observer in the laboratory reference frame. This
similarity is of course not complete since particles in the
described co-moving frame should experience inertia forces. However
the inertia forces are determined only by local geometric
characteristics of the frame. The locality of inertia forces and the
stationarity of the density allow us to consider a
small volume element in the co-moving frame as an independent
many-body system. Therefore we can 
extend the derivation of the static LDA to the time-dependent case.

The description of a flow in terms of trajectories of small liquid
element is the main idea behind the Lagrangian formulation of the classical
continuum mechanics \cite{PhysAc}. One can show that the
transformation to the Lagrangian coordinates exactly corresponds to the
transformation to the co-moving reference frame. In the next
section we apply the general description of quantum many-body dynamics in
the Lagrangian frame \cite{LagrFrameI} to the corresponding reformulation
of TDDFT.

\section{Many-body theory in the Lagrangian frame and geometric
  formulation of TDDFT}

\subsection{Many-body problem in the co-moving frame}
First we briefly review the key results of the many body-theory in
the Lagrangian frame (all details and derivations can be found in I).
The co-moving Lagrangian reference frame is defined as follows. Let 
${\bf v}({\bf x},t)={\bf j}({\bf x},t)/n({\bf x},t)$ be the velocity
of the flow. By solving the following initial value problem
\begin{equation}
\frac{\partial {\bf x}({\bm \xi},t)}{\partial t} =
{\bf v}({\bf x}({\bm \xi},t),t), \quad {\bf x}({\bm \xi},0)={\bm \xi}
\label{24}
\end{equation}
we find the function ${\bf x}({\bm \xi},t)$, which describes the
trajectory of a fluid element. The initial point, $\bm\xi$, of the
trajectory can be used as a unique label of the element. This initial
position of an infinitesimal fluid element is called the Lagrangian
coordinate. The transformation from the original ${\bf x}$-space to
the $\bm\xi$-space of initial positions is the transformation from the
Eulerian to the Lagrangian description of a fluid \cite{PhysAc}. On
the other hand, the equation ${\bf x}={\bf x}({\bm \xi},t)$, which
maps ${\bf x}$ to $\bm\xi$, exactly corresponds to the transformation
to the frame, attached to a fluid element. One of the most important
characteristics of the Lagrangian frame is Green's deformation tensor
\cite{PhysAc}, $g_{\mu\nu}({\bm \xi},t)$,
\begin{equation}
g_{\mu\nu}= \frac{\partial x^{\alpha}}{\partial \xi^{\mu}}
            \frac{\partial x^{\alpha}}{\partial \xi^{\nu}},
\qquad
g^{\mu\nu}= \frac{\partial \xi^{\mu}}{\partial x^{\alpha}}
            \frac{\partial \xi^{\nu}}{\partial x^{\alpha}}
\label{25}
\end{equation}
Tensor $g_{\mu\nu}$ plays a role of metric in the Lagrangian
$\bm\xi$-space (we will use the notation $g$ for the determinant of
$g_{\mu\nu}$). It has been shown in I that the field
operators, $\widetilde{\psi}(\bm\xi,t)$, in the Lagrangian frame are related
to the field operators, $\psi({\bf x},t)$, in the laboratory frame as
follows 
$$
\widetilde{\psi}(\bm\xi,t)= 
g^{\frac{1}{4}}\psi({\bf x}({\bm\xi},t),t).
$$
Apparently the operators $\widetilde{\psi}(\bm\xi,t)$ satisfy the
common equal-time commutation relations, which is guaranteed by the factor
$g^{\frac{1}{4}}$ in their definition. The current operator,  
$\widehat{\widetilde{j}^{\mu}}(\bm\xi,t)$, and the
density operator, $\widehat{\widetilde{n}}(\bm\xi,t)$, in the
Lagrangian frame are defined by the following expressions 
\begin{eqnarray}
\widehat{\widetilde{n}}(\bm\xi,t)
&=& \widetilde{\psi}^{\dag}(\bm\xi,t)\widetilde{\psi}(\bm\xi,t),
\label{26}\\
\widehat{\widetilde{j}^{\mu}}(\bm\xi,t)
&=& g^{\mu\nu}\left[\frac{-i}{2m}\left(
 \widetilde{\psi}^{\dag}
\frac{\partial\widetilde{\psi}}{\partial\xi^{\nu}} -
\frac{\partial\widetilde{\psi}^{\dag}}{\partial\xi^{\nu}}
\widetilde{\psi}\right) -
\widetilde{v}_{\nu}\widetilde{\psi}^{\dag}\widetilde{\psi}
 \right]
\label{27}
\end{eqnarray}
where $\widetilde{v}_{\nu}=g_{\nu\mu}\widetilde{v}^{\mu}$ is the
covariant component of the velocity vector ${\bf v}$, transformed to the new
frame
$$
\widetilde{v}^{\mu}({\bm\xi},t)=
\frac{\partial \xi^{\mu}}{\partial x^{\nu}}
v^{\nu}({\bf x}({\bm\xi},t),t)
$$ 
The Heisenberg equation of motion for the density operator of
Eq.~(\ref{26}) takes a form of the operator continuity equation
\begin{equation}
\frac{\partial \widehat{\widetilde{n}}}{\partial t} +
\frac{\partial \widehat{\widetilde{j}^{\mu}}}{\partial \xi^{\mu}} = 0
\label{28}
\end{equation}
On the level of expectation values Eq.~(\ref{28}) is trivially
satisfied. One can check by the explicit calculations that the
expectation value of the current operator, Eq.~(\ref{27}), is zero, while
the expectation value of the density operator, Eq.~(\ref{26}), is
time-independent
\begin{eqnarray}
\widetilde{j}^{\mu}(\bm\xi,t) &=& \langle
\widehat{\widetilde{j}^{\mu}}(\bm\xi,t)\rangle = 0.
\label{29}\\
\widetilde{n}(\bm\xi,t) &=& 
\langle\widehat{\widetilde{n}}(\bm\xi,t)\rangle =
\widetilde{n}(\bm\xi,0) = n_{0}(\bm\xi),
\label{30}
\end{eqnarray}
where $n_{0}({\bf x})$ is the initial density distribution. Equations
(\ref{29}) and (\ref{30}) are in complete agreement with the
qualitative discussion in the previous section. 

According to the results of the paper I, the vector
$m\widetilde{v}_{\mu}(\bm\xi,t)$ plays a role of an effective 
vector potential in the equation of motion for
the field operator, $\widetilde{\psi}$.
In general the velocity vector $\widetilde{v}_{\mu}(\bm\xi,t)$ has
both potential 
(longitudinal) and rotational (transverse) parts. The potential
part of a vector potential can be always removed from the kinetic
energy operator by the gauge transformation. Therefore it is
convenient to separate explicitly the potential part, 
$\widetilde{v}_{L\mu} = \partial_{\mu} \varphi$, of the vector
$\widetilde{v}_{\mu}$ 
\begin{equation}
\widetilde{v}_{\mu} = \frac{\partial \varphi}{\partial \xi^{\mu}} 
+ \widetilde{v}_{T\mu},
\label{31}
\end{equation}
where $\widetilde{v}_{T\mu}$ is the transverse part of
$\widetilde{v}_{\mu}$. Performing the gauge transformation 
$\widetilde{\psi}= e^{im\varphi}\widetilde{\psi}'$ in the equation of
motion for $\widetilde{\psi}$ (see Eq.~(34) in I) we obtain the
following equation of motion for the transformed operator 
$\widetilde{\psi}'$
\begin{widetext}
\begin{equation}
i\frac{\partial \widetilde{\psi}'({\bm\xi})}{\partial t} = 
g^{-\frac{1}{4}}\frac{
\widehat{K}_{\mu}\sqrt{g}\widehat{K}^{\mu}}{2m}
g^{-\frac{1}{4}}\widetilde{\psi}'({\bm\xi})
+ \int d{\bm\xi'}w(l_{\bm\xi,\bm\xi'})
\Delta\widehat{\widetilde{n}}({\bm\xi'})\widetilde{\psi}'({\bm\xi})
+ \left(m\frac{\partial\varphi}{\partial t} + U -
m\frac{\widetilde{v}_{\mu}\widetilde{v}^{\mu}}{2}\right)
\widetilde{\psi}'({\bm\xi}),
\label{32}
\end{equation}
\end{widetext}
where $\Delta\widehat{\widetilde{n}}({\bm\xi},t) =
\widehat{\widetilde{n}}({\bm\xi},t) - \widetilde{n}({\bm\xi},t)$ (the
Hartree term is included in $U=U_{\text{ext}}+U_{\text{H}}$).
Other notations in Eq.~(\ref{32}) are the same as in I: 
$$
 \widehat{K}_{\mu} =
  -i\frac{\partial}{\partial\xi^{\mu}} - m\widetilde{v}_{T\mu}
$$ 
is the operator of kinematic momentum, and $l_{\bm\xi,\bm\xi'}$ is the
length of geodesic connecting points $\bm\xi$ and $\bm\xi'$ 
(everywhere rising and lovering of tensor indexes are performed
according to the standard rules, i.e. $A_{\mu}=g_{\mu\nu}A^{\nu}$,
etc.). The deformation tensor and the velocity vector in Eq.~(\ref{32})
describe generalized inertial forces in the local noninertial
reference frame. Tensor $g_{\mu\nu}$ in the kinetic energy term
produces the ``geodesic'' force. This inertia force is responsible for
the motion of a free particle along the geodesic in $\bm\xi$-space. The
velocity $\widetilde{v}_{\mu}$, which acts as a vector potential in
Eq.~(\ref{32}), produces the Coriolis 
force (an effective Lorentz force) and the linear acceleration force
(an effective electric field). The last term in the brackets in
Eq.~(\ref{32}) is responsible for the inertia force that is related to
the kinetic energy of the frame (an analog of the centrifugal force). 

Equation (\ref{32}) is the equation of motion in a reference frame
moving with some velocity ${\bf v}$. In fact, the form of
Eq.~(\ref{32}) is covariant under an arbitrary transformation of
coordinates, which is generated by a continuous vector valued function ${\bf
v}({\bf x},t)$ via Eq.~(\ref{24}). To specify a particular reference
frame we need to impose an additional ``gauge'' condition. The gauge
condition assigns a particular value to the generating function ${\bf
v}({\bf x},t)$. There are a few formal possibilities to specify the
co-moving Lagrangian frame (see I). For example, since the expectation
value of the current operator in the Lagrangian frame should be zero,
we can impose the condition of Eq.~(\ref{29}) on the solutions to the
equation of motion, Eq.~(\ref{32}). In the present paper we prefer to
use another gauge fixing condition. Namely we require that the
solution to the equation of motion, Eq.~(\ref{32}), should be
consistent with the equation of the local force balance in the 
Lagrangian frame
\begin{equation}
m\frac{\partial\widetilde{v}_{T\mu}}{\partial t}
+ \frac{\partial}{\partial\xi^{\mu}}\left(
m\frac{\partial\varphi}{\partial t} + U
- m\frac{\widetilde{v}_{\nu}\widetilde{v}^{\nu}}{2}\right)
+ \frac{\sqrt{g}}{n_{0}}\widetilde{P}^{\nu}_{\mu ; \nu} = 0
\label{33}
\end{equation}
where $\widetilde{P}^{\nu}_{\mu ; \nu}$ is the covariant divergence
\cite{DubrovinI,LandauII:e} of the stress tensor 
\begin{equation}
\widetilde{P}^{\nu}_{\mu ; \nu} =  \frac{1}{\sqrt{g}}
\frac{\partial\sqrt{g}\widetilde{P}^{\nu}_{\mu}}{\partial\xi^{\nu}}
- \frac{1}{2}\frac{\partial g_{\alpha\beta}}{\partial\xi^{\mu}}
\widetilde{P}^{\alpha\beta}
\label{34}
\end{equation}
The force balance equation of Eq.~(\ref{33}) corresponds to
the local momentum conservation law, Eq.~(\ref{11}), transformed to
the Lagrangian frame. The stress tensor in the Lagrangian frame, 
$\widetilde{P}_{\mu\nu}$, which enters Eq.~(\ref{33}), is a linear
functional of the one particle density matrix $\widetilde{\rho}_{1}$,
and of the pair correlation function $\widetilde{G}_{2}$:
\begin{equation}
\widetilde{P}_{\mu\nu} = 
\widetilde{P}_{\mu\nu}[\widetilde{\rho}_{1},\widetilde{G}_{2}](\bm\xi,t)
\label{35}
\end{equation}
The explicit microscopic form of the functional
$\widetilde{P}_{\mu\nu}[\widetilde{\rho}_{1},\widetilde{G}_{2}]$ is
presented in Appendix~A (see Eqs.~(\ref{A1}), (\ref{A4}) and (\ref{A5})). 

Equation (\ref{33}) has precisely the same
physical significance as the static force balance equation of
Eq.~(\ref{15}). It shows that the inertia forces exactly compensate
the external force, $\frac{\partial}{\partial\xi^{\mu}}U$, and the
force of internal stress, 
$\frac{\sqrt{g}}{n_{0}}\widetilde{P}^{\nu}_{\mu ; \nu}$. The result of
this compensation is the absence of the current, and the stationary
density distribution in the Lagrangian frame.

Equations (\ref{32}) and (\ref{33}) constitute the full set of equations
of quantum many-body theory in the Lagrangian frame.

\subsection{TDDFT in the Lagrangian frame. Stress tensor as a
  universal functional of the dynamic deformation}

Now we are ready to the discussion of TDDFT. The complete description of
many-body dynamics in the Lagrangian frame corresponds to the solution of
the equation of motion, Eq.~(\ref{32}), supplemented by the frame
fixing condition of Eq.~(\ref{33}). Let us note that both
Eq.~(\ref{32}) and Eq.~(\ref{33}) contain the same effective potential
(the term in the brackets in Eq.~(\ref{32}) and Eq.~(\ref{33})). Using
this simple property we can 
formulate the following two-step procedure for solving the system of
Eqs.~(\ref{32}), (\ref{33}). On the first step we solve a
universal nonlinear many-body problem of the form
\begin{eqnarray}\nonumber
i\frac{\partial \widetilde{\psi}'({\bm\xi})}{\partial t} &=& 
g^{-\frac{1}{4}}\frac{
\widehat{K}_{\mu}\sqrt{g}\widehat{K}^{\mu}}{2m}
g^{-\frac{1}{4}}\widetilde{\psi}'({\bm\xi}) \\
&+& \left[\int d{\bm\xi'}w(l_{\bm\xi,\bm\xi'})
\Delta\widehat{\widetilde{n}}({\bm\xi'})
+ U_{\text{s-c}}(\bm\xi,t)\right]\widetilde{\psi}'({\bm\xi})\quad
\label{36}
\end{eqnarray}
where the effective potential $U_{\text{s-c}}(\bm\xi,t)$ is the
solution to the following self-consistency equation 
\begin{equation}
-\frac{\partial}{\partial\xi^{\mu}}U_{\text{s-c}}(\bm\xi,t)
= \frac{\sqrt{g}}{n_{0}}
\widetilde{P}^{\nu}_{\mu ; \nu}[\widetilde{\rho}_{1},\widetilde{G}_{2}]
+ m\frac{\partial\widetilde{v}_{T\mu}}{\partial t}
\label{37}
\end{equation}
The initial conditions for Eqs.~(\ref{36}) and (\ref{37}) are the same as
in the original physical many
body-problem, Eqs.~(\ref{32}), (\ref{33}). The special form of the
self-consistency equation, Eq.~(\ref{37}), ensures the stationarity
of the particles density and zero current density. Indeed, using
Eqs.~(\ref{36}) and (\ref{37}) we obtain the following
equations of motion for the density $\widetilde{n}$ and for the
current $\widetilde{\bf j}$
\begin{eqnarray}
\frac{\partial \widetilde{n}}{\partial t} &+&
\frac{\partial \widetilde{j}^{\mu}}{\partial \xi^{\mu}} = 0
\label{37a}\\
\frac{\partial \widetilde{j}_{\mu}}{\partial t} &+&
\widetilde{F}_{\mu\nu}\widetilde{j}_{\nu} = 0
\label{37b}
\end{eqnarray}
where $\widetilde{F}_{\mu\nu}$ is the skew-symmetric vorticity tensor,
which plays a role of an effective magnetic field
\begin{equation}
\widetilde{F}_{\mu\nu} =
\frac{\partial\widetilde{v}_{T\mu}}{\partial\xi^{\nu}} -
\frac{\partial\widetilde{v}_{T\nu}}{\partial\xi^{\mu}} =
\frac{\partial\widetilde{v}_{\mu}}{\partial\xi^{\nu}} -
\frac{\partial\widetilde{v}_{\nu}}{\partial\xi^{\mu}},
\label{38}
\end{equation}
The cancellation of the ``external'' force $\nabla U_{\text{s-c}}$, and the
inertial and the stress forces in Eq.~(\ref{37b}) is a consequence
of the self-consistency equation, Eq.~(\ref{37}). By solving
Eqs.~(\ref{37a}), (\ref{37b}) with the initial conditions  $\widetilde{\bf
  j}(\bm\xi,0)=0$, and $\widetilde{n}(\bm\xi,0)=n_{0}(\bm\xi)$, we
indeed confirm that for all $t>0$ $\widetilde{\bf j}(\bm\xi,t)=0$ and
$\widetilde{n}(\bm\xi,t)=n_{0}(\bm\xi)$. 

The self-consistent
nonlinear problem of Eqs.~(\ref{36}), (\ref{37}) is universal in that
sense that no external potential enters the equations. The
only ``external'' variables in Eqs.~(\ref{36}) and (\ref{37})  are the
deformation tensor $g_{\mu\nu}(\bm\xi,t)$ (an effective metric), and
the transverse part of the velocity, $\widetilde{v}_{T\mu}(\bm\xi,t)$,
(an effective vector potential). The vector
$\widetilde{v}_{T\mu}$ is uniquely determined by the skew-symmetric
vorticity tensor $\widetilde{F}_{\mu\nu}$ (an effective magnetic
field), Eq.~(\ref{38}). Therefore by 
solving the nonlinear problem of
Eqs.~(\ref{36}), (\ref{37}) we find the many-body density matrix as a
functional of two-basic variables, $g_{\mu\nu}(\bm\xi,t)$ and
$\widetilde{F}_{\mu\nu}(\bm\xi,t)$ \cite{note2}. Inserting this density
matrix into the 
microscopic expression for the stress tensor, Eq.~(\ref{35}), we obtain
the universal functional
\begin{equation}
\widetilde{P}_{\mu\nu} = 
\widetilde{P}_{\mu\nu}[g_{\mu\nu},\widetilde{F}_{\mu\nu}](\bm\xi,t).
\label{39}
\end{equation}
Calculation of the stress tensor functional, Eq.~(\ref{39}) completes
the first step in the solution of the original many-body
problem. 

The symmetric Green's
deformation tensor, $g_{\mu\nu}$, and the skew-symmetric vorticity
tensor, $\widetilde{F}_{\mu\nu}$, completely characterize the deformed
state of a fluid in the Lagrangian description. Therefore Eq.~(\ref{39})
can be interpreted as the exact nonequilibrium ``equation of state''
that relates the stress tensor to the dynamic deformation in the
system. Since $g_{\mu\nu}$ and $\widetilde{F}_{\mu\nu}$ are the
functionals of velocity, the stress tensor of Eq.~(\ref{39}) is also a
functional of velocity in agreement with Runge-Gross theorem. However,
the present interpretation of $\widetilde{P}_{\mu\nu}$ as a
deformation functional looks more natural physically. 

Substituting the ``equation of state'',
Eq.~(\ref{39}), into the force balance equation of Eq.~(\ref{33})
we get the exact quantum ``Navier-Stokes'' equation in the Lagrangian
formulation. The full set of the exact hydrodynamics equations
includes Eq.~(\ref{33}) and the trajectory equation,
Eq.~(\ref{24}). The solution of the system of Eqs.~(\ref{24}),
(\ref{33}) corresponds to the second step in the solution of the original
many-body problem. On this step we determine the evolution of velocity
for a given external 
potential. Equations (\ref{24}) and (\ref{33}) with the stress tensor
of Eq.~(\ref{39}) correspond to the exact TDDFT
hydrodynamics in the Lagrangian formulation of continuum mechanics.

The KS formulation of TDDFT requires a knowledge of the xc
potential $U_{\text{xc}}$. In Sec.~II we have shown that
$U_{\text{xc}}$ is related to the xc stress tensor 
$\widetilde{P}^{\text{xc}}_{\mu\nu} = \widetilde{P}_{\mu\nu} - 
\widetilde{T}^{\text{S}}_{\mu\nu}$, where
$\widetilde{T}^{\text{S}}_{\mu\nu}$ is the stress tensor for the
noninteracting KS 
system. Obviously, the KS stress tensor can be found from the solution
of a nonlinear noninteracting problem that corresponds to
Eqs.~(\ref{36}), (\ref{37}) with $w(l_{\bm\xi,\bm\xi'})=0$. Hence by
solving Eqs.~(\ref{36}), (\ref{37}) with and without interaction we
compute $\widetilde{P}_{\mu\nu}$ and $\widetilde{T}^{\text{S}}_{\mu\nu}$
respectively. The difference of these tensors gives us the xc stress
tensor in the Lagrangian frame as a functional of  $g_{\mu\nu}$ and 
$\widetilde{F}_{\mu\nu}$ 
\begin{equation}
\widetilde{P}^{\text{xc}}_{\mu\nu} = 
\widetilde{P}^{\text{xc}}_{\mu\nu}
[g_{\mu\nu},\widetilde{F}_{\mu\nu}](\bm\xi,t).
\label{40}
\end{equation}
Transforming the xc stress tensor of Eq.~(\ref{40}) back to the
laboratory frame, and substituting the result into Eqs.~(\ref{20}),
(\ref{21}), we obtain the equation for the xc potential
$U_{\text{xc}}({\bf x},t)$. Another possibility is to determine xc
potential, $\widetilde{U}_{\text{xc}}(\bm\xi,t)$, directly in the Lagrangian
frame by solving the following  equation
\begin{equation}
\frac{\partial}{\partial\xi^{\mu}}\widetilde{U}_{\text{xc}}(\bm\xi,t)
= \frac{\sqrt{g}}{n_{0}}
\widetilde{P}^{\text{xc}\nu}_{\mu ; \nu}
\label{41}
\end{equation}
The transformation of $\widetilde{U}_{\text{xc}}(\bm\xi,t)$ to the
laboratory frame corresponds to the following replacement 
$\bm\xi\to\bm\xi({\bf x},t)$, i.e.
\begin{equation}
U_{\text{xc}}({\bf x},t)=\widetilde{U}_{\text{xc}}
[g_{\mu\nu},\widetilde{F}_{\mu\nu}](\bm\xi({\bf x},t),t)
\label{42}
\end{equation}
where $\widetilde{U}_{\text{xc}}$ is the solution to Eq.~(\ref{41}). 
  
Let us note that the problem of calculation of the equilibrium stress
tensor functional, 
$P_{\mu\nu}$, in the static DFT can be formulated in
exactly the same fashion. To calculate $P_{\mu\nu}[n]({\bf x})$ we
need to find the equilibrium solution to the following universal nonlinear
many-body problem
\begin{eqnarray}\nonumber
i\frac{\partial}{\partial t}\psi({\bf x}) &=&
-\frac{\nabla^{2}}{2m}\psi({\bf x})
+ U_{\text{s-c}}({\bf x})\psi({\bf x})\\
&+&\int d{\bf x'}w(|{\bf x-x'}|)
\Delta\widehat{n}({\bf x'})\psi({\bf x})
\label{43}\\
\frac{\partial}{\partial x^{\mu}} U_{\text{s-c}}({\bf x}) &=& 
\frac{1}{n}\frac{\partial}{\partial x^{\nu}} P_{\mu\nu}[\rho_{1},G_{2}],
\label{44}
\end{eqnarray}
where 
$P_{\mu\nu}[\rho_{1},G_{2}]=T_{\mu\nu}[\rho_{1}]+W_{\mu\nu}[G_{2}]$ is
defined after Eqs.~(\ref{13}), (\ref{14}). For a given density $n({\bf x})$
the equilibrium solution to Eqs.~(\ref{43}), (\ref{44}) defines the
stress tensor $P_{\mu\nu}$ as a universal functional of $n$.   

Stationarity of the density in the Lagrangian frame makes the dynamic
universal problem of Eqs.~(\ref{36}), (\ref{37}) to a large
extent similar to the equilibrium universal problem of
Eqs.~(\ref{43}), (\ref{44}). In Sec.~V we use this
similarity to derive a local nonadiabatic approximation in TDDFT.  

\section{Time-dependent local deformation approximation}

In the previous section we have shown that the 
calculation of xc stress tensor, which defines the xc potential,
reduces to the solution of the nonlinear universal many-body problem,
Eqs.~(\ref{36}), (\ref{37}). Obviously, it is not possible to solve
this problem exactly. However one can try to find an approximate
solution by a perturbative expansion in terms of some small
parameter. Below we construct a local approximation that corresponds to
the lowest order in the gradients of basic variables (i.e. the density in
the static DFT, and the deformation tensor in TDDFT).

\subsection{General formulation of a nonadiabatic local approximation}

\subsubsection{Preliminaries: Derivation of the static LDA}
To illustrate the general procedure we start again with the familiar
case of the equilibrium theory. The problem is to find the equilibrium
solution to Eqs.~(\ref{43}), (\ref{44}), assuming that gradients of the
density $n$ are vanishingly small. In the limit $\nabla n\to 0$ the
spatial derivatives of the stress tensor also vanish. Hence to the
lowest order in the density gradients the solution to the
self-consistency Eq.~(\ref{44}) takes a trivial form, $U_{\text{s-c}}({\bf
  x})=C$, where $C$ is a constant. Therefore the many-body equation of
motion, Eq.~(\ref{43}), simplifies as follows
\begin{equation}
i\frac{\partial\psi({\bf x})}{\partial t} =
-\frac{\nabla^{2}}{2m}\psi({\bf x})
+ \int d{\bf x'}w(|{\bf x-x'}|)
\widehat{n}({\bf x'})\psi({\bf x}).
\label{45}
\end{equation}
Thus the nonlinear problem of Eqs.~(\ref{43}), (\ref{44}) reduces to
the usual linear many-body problem for a homogeneous equilibrium system
with a given density $n$. Substituting the equilibrium solution to
Eq.~(\ref{45}) into Eqs.~(\ref{13a})--(\ref{14}) we compute the stress
tensor, $P_{\mu\nu}(n)=\delta_{\mu\nu}P(n)$, where $P$ is the pressure of
the homogeneous system: 
\begin{equation}         
P(n) = \frac{2}{d}E_{\text{kin}} - \frac{1}{2d}\int {\bf x}
\frac{\partial w(|{\bf x}|)}{\partial{\bf x}}
G_{2}^{\text{eq}}(|{\bf x}|)d{\bf x}.
\label{46}
\end{equation}
Here $d$ is the number of spatial dimensions, $E_{\text{kin}}$ is the
kinetic energy per unit volume, and $G_{2}^{\text{eq}}(|{\bf x}|)$ is
the pair correlation function of the equilibrium homogeneous
system. Similarly by solving the homogeneous noninteracting problem we
find the KS stress tensor
$T^{\text{S}}_{\mu\nu}(n)=\delta_{\mu\nu}\frac{2}{d}E_{\text{kin}}^{(0)}$,
where $E_{\text{kin}}^{(0)}$ is the kinetic energy of 
an ideal Fermi gas. Substituting $P_{\mu\nu}(n({\bf x}))$ and 
$T^{\text{S}}_{\mu\nu}(n({\bf x}))$ into Eqs.~(\ref{19}) and
(\ref{18}) we recover the common static LDA (see Sec.~III).

\subsubsection{Basic equations of TDLDA: The homogeneous many-body problem}
The above procedure allows for a straightforward extension to the
time-dependent problem. Let us assume that the characteristic length
scale, $L$, of the deformation inhomogeneity goes to infinity. In this
limit the vector $\frac{\sqrt{g}}{n_{0}}\widetilde{P}^{\nu}_{\mu ; \nu}$
in the right hand side in Eq.~(\ref{37}) vanishes. Therefore to the
lowest order in $1/L\to 0$ the self-consistent solution to
Eq.~(\ref{37}) takes the form: $U_{\text{s-c}}({\bf x},t)=C(t)$ and
${\bf v}_{T}=0$. Substituting this solution into Eq.~(\ref{36}) we
get the equation of motion for $\widetilde{\psi}$-operator
\begin{equation}
i\frac{\partial\widetilde{\psi}(\bm\xi)}{\partial t} =
- \frac{g_{\mu\nu}(t)}{2m}
\frac{\partial^{2}\widetilde{\psi}(\bm\xi)}{\partial\xi^{\mu}\partial\xi^{\nu}}
+ \int d\bm\xi'w(\|\bm\xi-\bm\xi'\|)
\widehat{\widetilde{n}}({\bm\xi'})
\widetilde{\psi}(\bm\xi)
\label{47}
\end{equation}
where $\|\bm\xi-\bm\xi'\|=l_{\bm\xi,\bm\xi'}$ is the length of
geodesic in a homogeneously deformed Lagrangian space (see Appendix~B):
\begin{equation}
\|\bm\xi-\bm\xi'\| = 
\sqrt{g_{\mu\nu}(t)(\xi^{\mu}-\xi'^{\mu})(\xi^{\nu}-\xi'^{\nu})}.
\label{48}
\end{equation}

Equation (\ref{47}) corresponds to a homogeneous many-body
system. It is more natural to reformulate this homogeneous problem using
the momentum representation for field operators
\begin{equation}
\widetilde{\psi}(\bm\xi)=
\sum_{\bf k}e^{ik_{\mu}\xi^{\mu}}\widetilde{a}_{\bf k},
\label{49}
\end{equation}
The equation of motion for annihilation operator, 
$\widetilde{a}_{\bf k}$, takes the form 
\begin{equation}
i\frac{\partial \widetilde{a}_{\bf k}}{\partial t} = 
g^{\mu\nu}(t)\frac{k_{\mu}k_{\nu}}{2m}\widetilde{a}_{\bf k} +
\sum_{\bf p,q}\frac{\bar{w}(\|{\bf q}\|)}{\sqrt{g(t)}}
\widetilde{a}^{\dag}_{\bf p}
\widetilde{a}_{\bf p+q}\widetilde{a}_{\bf k-q}
\label{50}
\end{equation} 
where $\bar{w}(q)$ is the Fourier component of the interaction
potential, and
\begin{equation} 
\|{\bf q}\| =
\sqrt{g^{\alpha\beta}(t)q_{\beta}q_{\alpha}}
\label{51}
\end{equation}
is the norm of the wave vector in the deformed momentum space.
Equation~(\ref{50}), corresponds to the following Hamiltonian
\begin{equation}
\widetilde{H} = \sum_{\bf k}g^{\mu\nu}\frac{k_{\mu}k_{\nu}}{2m}
\widetilde{a}^{\dag}_{\bf k}\widetilde{a}_{\bf k} +
\frac{1}{2\sqrt{g}}\sum_{\bf k,q}\bar{w}(\|{\bf q}\|)
\widetilde{a}^{\dag}_{\bf k}
\widehat{\widetilde{n}}_{\bf q}\widetilde{a}_{\bf k-q}
\label{52}
\end{equation} 
where $\widehat{\widetilde{n}}_{\bf q}=\sum_{\bf p}
\widetilde{a}^{\dag}_{\bf p}\widetilde{a}_{\bf p+q}$ is the density
operator in the momentum representation.  
In Appendix B we show that to the lowest order in the
deformation gradients the microscopic expression for the
stress tensor, $\widetilde{P}_{\mu\nu}$, simplify as follows
\begin{eqnarray}\nonumber
\widetilde{P}_{\mu\nu} &=& \frac{1}{\sqrt{g}}\sum_{\bf k}
\frac{k_{\mu}k_{\nu}}{m}\widetilde{f}({\bf k})\\
&+& \frac{1}{2g}\sum_{\bf k}\left[
\frac{k_{\mu}k_{\nu}}{\|{\bf k}\|}\bar{w}'(\|{\bf k}\|)
+ g_{\mu\nu}\bar{w}(\|{\bf k}\|)\right]\widetilde{G}_{2}({\bf k}),\quad
\label{53}
\end{eqnarray}
where $\widetilde{f}({\bf k})=
\langle\widetilde{a}^{\dag}_{\bf k}\widetilde{a}_{\bf k}\rangle$ is
the Wigner function, $\widetilde{G}_{2}({\bf k})$ is the Fourier
component of the pair correlation function, and 
$\bar{w}'(x)=d\bar{w}(x)/dx$. 

It should be mentioned that Eq.~(\ref{53}) can be derived directly
from the ``geometric'' definition of the stress tensor (see I and
Ref.~\onlinecite{RogRap2002})
$$ 
\widetilde{P}_{\mu\nu} = \frac{2}{\sqrt{g}}
\left\langle
\frac{\delta \widetilde{H}}{\delta g^{\mu\nu}}
\right\rangle.
$$
Indeed, using the relation $\delta g = - g g_{\mu\nu}\delta g^{\mu\nu}$,
and computing the derivative of the Hamiltonian, Eq.~(\ref{52}),
with respect to $g^{\mu\nu}$, we immediately recover Eq.~(\ref{53}).

The Hamiltonian of Eq.~(\ref{52}) determines the homogeneous problem
which we need to solve for the derivation of a local approximation in
TDDFT. This problem corresponds to a system of particles in a small
volume located at the point $\bm\xi$ of Lagrangian space. The density
of particles is time-independent and equals to the initial density,
$n_{0}(\bm\xi)$, at that point (obviously, the operator of the number
of particles commutes with $\widetilde{H}$). The behavior of the
system is governed by the local value of the deformation tensor,
$g_{\mu\nu}(\bm\xi,t)$. By solving the equations of motion we find the
Wigner function, $\widetilde{f}({\bf k},t)$, and the pair correlation
function, $\widetilde{G}_{2}({\bf k},t)$. Substitution of
$\widetilde{f}({\bf k},t)$ and $\widetilde{G}_{2}({\bf k},t)$ into
Eq.~(\ref{53}) yields the stress tensor functional
$\widetilde{P}_{\mu\nu}[g_{\mu\nu}(\bm\xi,t),n_{0}(\bm\xi)]$. By the
repetition of the above procedure for the noninteracting system
(Eqs.~(\ref{52}), (\ref{53}) with $\bar{w}=0$) we find the KS stress
tensor,
$\widetilde{T}^{\text{S}}_{\mu\nu}[g_{\mu\nu}(\bm\xi,t),n_{0}(\bm\xi)]$,
and, finally, the xc stress tensor
\begin{equation}  
\widetilde{P}^{\text{xc}}_{\mu\nu}[g_{\mu\nu}(\bm\xi,t),n_{0}(\bm\xi)]
= \widetilde{P}_{\mu\nu} - \widetilde{T}^{\text{S}}_{\mu\nu}.
\label{54}
\end{equation}
Substituting $\widetilde{P}^{\text{xc}}_{\mu\nu}$ of Eq.~(\ref{54})
into Eq.~(\ref{41}) we determine the corresponding xc potential in
the Lagrangian frame. 

The xc stress tensor $\widetilde{P}^{\text{xc}}_{\mu\nu}$,
Eq.~(\ref{54}), is a local in space functional of the deformation
tensor (it should be noted that in general this functional
is nonlocal in time). In what follows the approximation
of Eq.~(\ref{54}) will be referred to as a {\em Time-Dependent Local
Deformation Approximation} (TDLDA). 

The construction of TDLDA reduces to the solution of the homogeneous
many-body problem. In this respect the situation is similar to the static
case. However, the homogeneous time-dependent problem, defined by the
Hamiltonian of Eq.~(\ref{52}), is still too complicated to be solved
exactly. Indeed the operator equation of motion, Eq.~(\ref{50}),
generates an infinite set of coupled evolution equations (BBGKY
hierarchy \cite{Klimont:e}) for correlation functions. The first
equation of this hierarchy is the equation of motion for the Wigner
function
\begin{equation}
i\frac{\partial \widetilde{f}({\bf k})}{\partial t} = 
\sum_{\bf p,q}\frac{\bar{w}(\|{\bf q}\|)}{\sqrt{g}}\left\langle
\widetilde{a}^{\dag}_{\bf p}\left(
\widetilde{a}^{\dag}_{\bf k}\widetilde{a}_{\bf k-q}
- \widetilde{a}^{\dag}_{\bf k+q}\widetilde{a}_{\bf k}
\right)\widetilde{a}_{\bf p+q}\right\rangle
\label{55}
\end{equation}
An equation for the four-fermion correlator, entering the right
hand side in Eq.~(\ref{55}), couples to the six-fermion correlation functions,
etc. However the homogeneity of the problem and a very specific
form of the ``driving force'' in the equations of motion allow us to
construct reasonable approximate xc functionals (see next subsections).

\subsubsection{Stress tensor of the noninteracting KS system}

A necessary step in the derivation of TDLDA is to compute   
the stress tensor, $\widetilde{T}^{\text{S}}_{\mu\nu}$, in the
noninteracting system. This problem can be solved exactly. In the
noninteracting case ($\bar{w}=0$) Eqs.~(\ref{53}) and (\ref{55})
reduce to the following simple form    
\begin{eqnarray}
\widetilde{T}^{\text{S}}_{\mu\nu} &=& \frac{1}{\sqrt{g}}\sum_{\bf k}
\frac{k_{\mu}k_{\nu}}{m}\widetilde{f}({\bf k},t),
\label{56}\\
&&\frac{\partial}{\partial t}\widetilde{f}({\bf k},t) = 0.
\label{57}
\end{eqnarray}
Equation (\ref{57}) shows that the distribution function of
noninteracting particles in the Lagrangian frame is time-independent. Let
us assume for definiteness that the system evolves from the
equilibrium state, i.~e. $\widetilde{f}({\bf k},0)=n^{F}_{\bf k}$,
where $n^{F}_{\bf k}$ is the Fermi function. In this case the solution
to Eq.~(\ref{57}) takes the form
\begin{equation}
\widetilde{f}({\bf k},t) = \widetilde{f}({\bf k},0)=n^{F}_{\bf k}.
\label{58}
\end{equation}
Substituting Eq.~(\ref{58}) into Eq.~(\ref{56}) we get the kinetic
stress tensor of the KS system in the Lagrangian frame
\begin{equation}
\widetilde{T}^{\text{S}}_{\mu\nu}(\bm\xi,t) =
\frac{\delta_{\mu\nu}}{\sqrt{g(\bm\xi,t)}}P_{0}(n_{0}(\bm\xi)),
\label{59}
\end{equation}
where the function $P_{0}(n)=\frac{2}{d}E_{\text{kin}}^{(0)}(n)$ is the
equilibrium kinetic pressure of a noninteracting homogeneous Fermi gas. 

For the practical calculation of xc potential in TDLDA 
(see Eqs.~(\ref{18}), (\ref{19}) ) we need the stress tensor, 
$T^{\text{S}}_{\mu\nu}({\bf x},t)$, in the laboratory frame. Application
of the common tensor transformation rules \cite{DubrovinI},  
\begin{equation}
P_{\mu\nu}({\bf x},t) = 
\frac{\partial \xi^{\alpha}}{\partial x^{\mu}}
\frac{\partial \xi^{\beta}}{\partial x^{\nu}}
\widetilde{P}_{\alpha\beta}(\bm\xi({\bf x},t),t),
\label{60}
\end{equation}
to the stress tensor of Eq.~(\ref{59}) yields the result
\begin{equation}
T^{\text{S}}_{\mu\nu}({\bf x},t) = 
\bar{g}_{\mu\nu}({\bf x},t)\sqrt{\bar{g}({\bf x},t)}
P_{0}(n_{0}(\bm\xi({\bf x},t)),
\label{61}
\end{equation}
were $\bar{g}_{\mu\nu}({\bf x},t)$ is the Cauchy's deformation tensor
\cite{PhysAc} 
\begin{equation}
\bar{g}_{\mu\nu}({\bf x},t) = 
\frac{\partial \xi^{\alpha}}{\partial x^{\mu}}
\frac{\partial \xi^{\alpha}}{\partial x^{\nu}}.
\label{62}
\end{equation}
The determinant, $\bar{g}({\bf x},t)$, of Cauchy's deformation tensor,
Eq.~(\ref{62}), is related to the determinant, $g(\bm\xi,t)$, of
Green's deformation tensor, Eq.~(\ref{25}), as follows
\begin{equation}
\bar{g}({\bf x},t) = g^{-1}(\bm\xi({\bf x},t),t). 
\label{63}
\end{equation}
Equation (\ref{61}) clearly demonstrates an extreme nonlocality which
is related to the memory effects. The stress tensor,
$T^{\text{S}}_{\mu\nu}({\bf x},t)$, at a given point ${\bf x}$ depends on
the initial density at the point $\bm\xi({\bf x},t)$ that is the initial
position of a fluid element presently at ${\bf x}$. Let us show
that this dependence on the delayed position can be represented in a
local form. By definition of the Lagrangian coordinate, the
density, $n({\bf x},t)$, in the laboratory frame can be expressed in
terms of the initial density distribution (the density in the Lagrangian
frame):
\begin{equation}
n({\bf x},t) = \frac{n_{0}(\bm\xi({\bf x},t))}{\sqrt{g(\bm\xi({\bf x},t),t)}}. 
\label{64}
\end{equation}
Using the relation of Eq.~(\ref{63}) we can represent the nonlocal
quantity $n_{0}(\bm\xi({\bf x},t))$ in the following form
\begin{equation}
n_{0}(\bm\xi({\bf x},t)) = \frac{n({\bf x},t)}{\sqrt{\bar{g}({\bf x},t)}}. 
\label{65}
\end{equation}
Substituting Eq.~(\ref{65}) into Eq.~(\ref{61}) we obtain a completely
local representation for the KS kinetic stress tensor
\begin{equation}
T^{\text{S}}_{\mu\nu}({\bf x},t) = 
\bar{g}_{\mu\nu}({\bf x},t)\sqrt{\bar{g}({\bf x},t)}
P_{0}\left(\frac{n({\bf x},t)}{\sqrt{\bar{g}({\bf x},t)}}\right).
\label{66}
\end{equation}
The nonlocality of the stress tensor in the form of Eq.~(\ref{61}) is now
hidden in the space-time dependence of the function
$\bar{g}({\bf x},t)$.           

\subsection{Exchange-only TDLDA}

The most difficult part in the derivation of an explicit TDLDA is the
solution of the interacting problem defined by the Hamiltonian of
Eq.~(\ref{52}). In this subsection we find the exact solution of this
problem in 
the exchange approximation, which provides us with the x-only
TDLDA. In the x-only case the pair correlation function
$\widetilde{G}_{2}({\bf k},t)$ is completely determined by the one
particle distribution function $\widetilde{f}({\bf k},t)$
\begin{equation}  
\widetilde{G}_{2}({\bf k},t) = -  
\sum_{\bf p}\widetilde{f}({\bf k+p},t)\widetilde{f}({\bf p},t)
\label{67}
\end{equation}
Performing the mean field decoupling of the four-fermion terms in
Eq.~(\ref{55}) we find that the right hand side in this equation
vanishes. Therefore the equation of motion for the function
$\widetilde{f}({\bf k},t)$ takes the form 
\begin{equation}  
\frac{\partial}{\partial t}\widetilde{f}({\bf k},t) = 0.
\label{68}
\end{equation}
Equation (\ref{68}) coincides with the corresponding equation of
motion for the noninteracting system, Eq.~(\ref{57}). Hence in the
x-only approximation both the Wigner function and the pair correlation
function in the Lagrangian frame preserve their initial form
\begin{eqnarray}
&&\widetilde{f}({\bf k},t) = n_{\bf k}^{F},
\label{69}\\   
\widetilde{G}_{2}({\bf k},t) &=& G_{2}^{\text{x}}(n_{0};k) = 
- \sum_{\bf p} n_{\bf k+p}^{F}n_{\bf p}^{F}.
\label{70}
\end{eqnarray}
Here $k=|{\bf k}|=\sqrt{k_{\mu}k_{\mu}}$ is the usual modulus of ${\bf
k}$, and $G_{2}^{\text{x}}(n;k)$ is the exchange pair correlation
function in the equilibrium Fermi gas of the density $n$. Substituting
Eqs.~(\ref{69}) and (\ref{70}) into Eq.~(\ref{53}) we obtain the
following stress tensor in the interacting system
\begin{equation}
\widetilde{P}_{\mu\nu} =
\frac{\delta_{\mu\nu}}{\sqrt{g(\bm\xi,t)}}P_{0}(n_{0}(\bm\xi))
+ \widetilde{P}^{\text{x}}_{\mu\nu}(n_{0}(\bm\xi), g_{\mu\nu}(\bm\xi,t)).
\label{71}
\end{equation}
The first term in the right hand side in Eq.~(\ref{71}) is the kinetic
stress tensor of the noninteracting system, while the second term, 
$\widetilde{P}^{\text{x}}_{\mu\nu}$, corresponds to the exchange
contribution to the local stress density
\begin{equation}
\widetilde{P}^{\text{x}}_{\mu\nu} = \frac{1}{2g}\sum_{\bf k}\left[
\frac{k_{\mu}k_{\nu}}{\|{\bf k}\|}\bar{w}'(\|{\bf k}\|)
+ g_{\mu\nu}\bar{w}(\|{\bf k}\|)\right]
G_{2}^{\text{x}}(n_{0};k)
\label{72}
\end{equation}
Using the transformation rules of Eq.~(\ref{60}) we get the
following expression for the exchange stress tensor in the
laboratory frame
\begin{eqnarray}\nonumber
P^{\text{x}}_{\mu\nu}(n,\bar{g}_{\alpha\beta}) 
&=& \frac{\sqrt{\bar{g}}}{2}\sum_{\bf p}\left[
\frac{p_{\mu}p_{\nu}}{p}\bar{w}'_{p}
+ \delta_{\mu\nu}\bar{w}_{p}\right]\\
&\times& G_{2}^{\text{x}}\left(\frac{n}{\sqrt{\bar{g}}};
\sqrt{\bar{g}^{\alpha\beta}p_{\alpha}p_{\beta}}\right),
\label{73}
\end{eqnarray}
where we introduced a shortcut notation
$\bar{w}_{p}=\bar{w}(p)$. Equations (\ref{73}), (\ref{21}) and
(\ref{20}) uniquely determine the local potential $U_{\text{x}}({\bf
x},t)$ in x-only TDLDA. Apparently the exchange potential
$U_{\text{x}}({\bf x},t)$ is a local (both in space and in time)
functional of the density $n({\bf x},t)$ and Cauchy's deformation
tensor $\bar{g}_{\mu\nu}({\bf x},t)$. In the equilibrium system
($\bar{g}_{\mu\nu}=\delta_{\mu\nu}$) the potential, defined by
Eqs.~(\ref{73}), (\ref{21}) and (\ref{20}), reduces to that in the
common static local exchange approximation.

\subsection{Linear response TDLDA.}

In the linear response regime the deformation tensor, $g_{\mu\nu}$,
slightly deviates from the Kronecker symbol
\begin{equation}  
g^{\mu\nu}(\bm\xi,t)\approx \delta_{\mu\nu} 
+ \delta g^{\mu\nu}(\bm\xi,t), \quad 
\delta g^{\mu\nu} = -\frac{\partial u_{\mu}}{\partial\xi^{\nu}}
- \frac{\partial u_{\nu}}{\partial\xi^{\mu}}, 
\label{74}
\end{equation}
where ${\bf u} = {\bf x}-\bm\xi$ is the displacement vector, which is
assumed to be small. In the linearized theory the trajectory equation
of Eq.~(\ref{24}) reduces to the common linear relation of the velocity
to the displacement
\begin{equation}
\frac{\partial {\bf u}(\bm\xi , t)}{\partial t} = {\bf v}(\bm\xi , t).
\label{74a}
\end{equation} 
Substituting $g^{\mu\nu}$ of Eq.~(\ref{74}) into Eq.~(\ref{52}) and
keeping only linear in $\delta g^{\mu\nu}$ terms, we obtain the
following linearized Hamiltonian
\begin{equation}
\widetilde{H} = H + \widehat{P}_{\mu\nu}\delta g^{\mu\nu},
\label{75}
\end{equation}
where $H$ is the standard Hamiltonian for the homogeneous system, and
$\widehat{P}_{\mu\nu}$ is the stress tensor operator
\begin{equation}
\widehat{P}_{\mu\nu} = \sum_{\bf k}\Big[\frac{k_{\mu}k_{\nu}}{m}
\widetilde{a}^{\dag}_{\bf k}\widetilde{a}_{\bf k}
+ \frac{1}{2}\Big(
\frac{k_{\mu}k_{\nu}}{k}\bar{w}'_{k}
+\delta_{\mu\nu}\bar{w}_{k}\Big)
\widehat{\widetilde{n}}_{\bf k}\widehat{\widetilde{n}}_{\bf -k}\Big]
\label{76}
\end{equation} 
First we need to compute the stress tensor $\widetilde{P}_{\mu\nu}$ in
the Lagrangian frame, Eq.~(\ref{53}). In the linear regime
Eq.~(\ref{53}) takes the form
\begin{equation}
\widetilde{P}_{\mu\nu} = \delta_{\mu\nu} P(n_{0}) 
+ \widetilde{Q}_{\mu\nu\alpha\beta}(\omega)\delta g^{\alpha\beta}(\omega)
\label{77}
\end{equation}
where the linear response kernel, 
$\widetilde{Q}_{\mu\nu\alpha\beta}(\omega)$, 
can be represented as follows
\begin{equation}
\widetilde{Q}_{\mu\nu\alpha\beta}(\omega) = 
\widetilde{Q}_{\mu\nu\alpha\beta}^{\infty} +
\Delta \widetilde{Q}_{\mu\nu\alpha\beta}(\omega).
\label{78}
\end{equation}
The first frequency independent term,
$\widetilde{Q}_{\mu\nu\alpha\beta}^{\infty}$, in Eq.~(\ref{78}) comes
from the explicit local in time dependence of the integrals in
Eq.~(\ref{53}) on the deformation tensor $g_{\mu\nu}(t)$. Namely, the
fourth-rank tensor $\widetilde{Q}_{\mu\nu\alpha\beta}^{\infty}$ is
defined by the following derivative
\begin{equation}  
\widetilde{Q}_{\mu\nu\alpha\beta}^{\infty} =
\left(\frac{\partial
\widetilde{P}_{\mu\nu}[f^{\text{eq}},G_{2}^{\text{eq}}]
}{\partial g^{\alpha\beta}}\right)_{g_{\mu\nu=\delta_{\mu\nu}}}
\label{79}
\end{equation}
where $\widetilde{P}_{\mu\nu}[f^{\text{eq}},G_{2}^{\text{eq}}]$ is the
stress tensor of Eq.~(\ref{53}), calculated with the equilibrium Wigner
function, $f^{\text{eq}}({\bf k})$, and the
equilibrium pair correlation function, $G_{2}^{\text{eq}}({\bf
  k})$. The perturbation (the second term) in the linearized Hamiltonian of
Eq.~(\ref{75}) induces deviations of the Wigner function and the pair
correlation function from their equilibrium values. These deviations
are responsible for the second, nonlocal in time term in Eq.~(\ref{78})
\begin{eqnarray}\nonumber
\Delta \widetilde{Q}_{\mu\nu\alpha\beta}
&=& \sum_{\bf k}\frac{k_{\mu}k_{\nu}}{2m}
\frac{\delta \widetilde{f}({\bf k},t)}{\delta  g^{\alpha\beta}(t')}\\
&+& \frac{1}{2}\sum_{\bf k}\Big(
\frac{k_{\mu}k_{\nu}}{k}\bar{w}'_{k}
+\delta_{\mu\nu}\bar{w}_{k}\Big)
\frac{\delta \widetilde{G}_{2}({\bf k},t)}{\delta  g^{\alpha\beta}(t')}.
\label{80}
\end{eqnarray} 
Comparing Eqs.~(\ref{75}), (\ref{76}) and (\ref{80}) we find that the
dynamic kernel $\Delta \widetilde{Q}_{\mu\nu\alpha\beta}$ can be
related to the following stress autocorrelation function
\begin{equation}  
\Delta\widetilde{Q}_{\mu\nu\alpha\beta}(\omega) =
- i \int_{0}^{\infty}dt e^{i\omega t}\left\langle
[\widehat{P}_{\mu\nu}(t),\widehat{P}_{\alpha\beta}(0)]
\right\rangle.
\label{81}
\end{equation}
Substituting the stress tensor in the Lagrangian frame,
Eq.~(\ref{77}), into the transformation formula of Eq.~(\ref{60}) we
compute the stress tensor in the laboratory frame
\begin{equation} 
P_{\mu\nu}({\bf x},\omega) = \delta_{\mu\nu}P(n_{0}({\bf x}))
+ \delta P_{\mu\nu}({\bf x},\omega), 
\label{82}
\end{equation}
where
\begin{equation} 
\delta P_{\mu\nu} = - \delta_{\mu\nu}\frac{\partial P}{\partial n_{0}}
{\bf u}\nabla n_{0} 
+ Q_{\mu\nu\alpha\beta}(\omega)g^{\alpha\beta}(\omega).
\label{83}
\end{equation}
The kernel $Q_{\mu\nu\alpha\beta}(\omega)$ in the laboratory frame
takes the form, which is similar to that of Eq.~(\ref{78})
\begin{equation}
Q_{\mu\nu\alpha\beta}(\omega) = 
Q_{\mu\nu\alpha\beta}^{\infty} +
\Delta \widetilde{Q}_{\mu\nu\alpha\beta}(\omega).
\label{84a}
\end{equation}
The frequency independent term,
$Q_{\mu\nu\alpha\beta}^{\infty}$, in Eq.~(\ref{84a}) is related to
the quantity $\widetilde{Q}_{\mu\nu\alpha\beta}^{\infty}$ 
of Eq.~(\ref{79}) as follows
\begin{equation} 
Q_{\mu\nu\alpha\beta}^{\infty} =
\frac{1}{2}P(\delta_{\mu\alpha}\delta_{\nu\beta}
+ \delta_{\nu\alpha}\delta_{\mu\beta})
+ \widetilde{Q}_{\mu\nu\alpha\beta}^{\infty}.
\label{84}
\end{equation}
It is worth mentioning that the first term in the right hand side in
Eq.~(\ref{83}) guarantees that the harmonic potential theorem
\cite{Dobson1994} is satisfied. In fact, only this term survives for
the rigid motion of the system. Within the present formalism this term
comes from the expansion of the argument of $\widetilde{P}_{\mu\nu}$
in the transformation rule of Eq.~(\ref{60}). The correction to the kernel
in the laboratory frame (the first term in Eq.~(\ref{84})) corresponds
to the expansion of the tensor prefactor in Eq.~(\ref{60}).

By symmetry the fourth-rank tensor $Q_{\mu\nu\alpha\beta}$ is uniquely
representable in the form
\begin{equation} 
Q_{\mu\nu\alpha\beta} = 
\left(\frac{K}{2} + \frac{\mu}{d}\right)
\delta_{\mu\nu}\delta_{\alpha\beta}
+\frac{\mu}{2}(\delta_{\mu\alpha}\delta_{\nu\beta}
+ \delta_{\nu\alpha}\delta_{\mu\beta}),
\label{85}
\end{equation}
The scalar coefficients,
$K(\omega)$ and $\mu(\omega)$, in Eq.~(\ref{85}) are related to the
tensor $Q_{\mu\nu\alpha\beta}$ contracted over different couples of
indexes
\begin{eqnarray}
K(\omega) &=& \frac{2}{d^{2}}Q_{\alpha\alpha\beta\beta}(\omega),
\label{86}\\
\mu(\omega) &=& \frac{2}{d^{2}+d-2}\left[
Q_{\alpha\beta\alpha\beta}(\omega) 
- \frac{1}{d}Q_{\alpha\alpha\beta\beta}(\omega)\right].
\label{87}
\end{eqnarray}
Substitution of Eq.~(\ref{85}) into Eq.~(\ref{83}) yields the
following result for the linear correction to the stress tensor in the
laboratory frame
\begin{eqnarray}\nonumber 
\delta P_{\mu\nu} = 
&-& \delta_{\mu\nu}\frac{\partial P}{\partial n_{0}}{\bf u}\nabla n_{0}\\ 
&+&\delta_{\mu\nu}K\frac{1}{2}\delta g^{\alpha\alpha}
+ \mu\left(\delta g^{\mu\nu} 
- \frac{\delta_{\mu\nu}}{d}\delta g^{\alpha\alpha}\right).
\label{88}
\end{eqnarray}  
The stress tensor $\delta P_{\mu\nu}$ of Eq.~(\ref{88}) has a clear
visco-elastic form, where $K(\omega)$ and $\mu(\omega)$ are the bulk
modulus and the shear modulus respectively. The xc stress tensor,
$\delta P_{\mu\nu}^{\text{xc}}$, is the difference of the expressions
given by Eq.~(\ref{88}) for the interacting and the noninteracting
systems. Apparently $\delta P_{\mu\nu}^{\text{xc}}$ takes a form of
Eq.~(\ref{88}) with $P$, $K$ and $\mu$ being replaced by
$P_{\text{xc}}$, $K_{\text{xc}}$ and $\mu_{\text{xc}}$ respectively,
where
\begin{eqnarray} 
P_{\text{xc}} &=& P - \frac{2}{d}E_{\text{kin}}^{(0)}, 
\label{89}\\
K_{\text{xc}} &=& K - K_{0}, \quad
\mu_{\text{xc}} = \mu - \mu_{0},
\label{90}
\end{eqnarray}  
are the xc pressure and the xc visco-elastic moduli. In Eqs.~(\ref{89}),
(\ref{90}) $E_{\text{kin}}^{(0)}$, $K_{0}$, and $\mu_{0}$ correspond to
the kinetic energy, the bulk modulus, and the shear modulus of an ideal
Fermi gas. Therefore in the linear
response regime our TDLDA naturally reduces to the Vignale-Kohn
approximation \cite{VigKohn1996} in the visco-elastic formulation of
Ref.~\onlinecite{VigUllCon1997}. 

An explicit microscopic representation for the bulk, $K$, and the
shear, $\mu$, moduli can be found using Eqs.~(\ref{87}), (\ref{86}),
(\ref{84}), (\ref{84a}), (\ref{81}) and (\ref{79}). Both $K$ and $\mu$
take the following general form
\begin{eqnarray} 
K(\omega) &=& K^{\infty} + \Delta K(\omega),
\label{91}\\
\mu(\omega) &=& \mu^{\infty} + \Delta \mu(\omega).
\label{92}
\end{eqnarray}  
The first terms, $K^{\infty}$ and $\mu^{\infty}$, in the right hand
sides in Eqs.~(\ref{91}) and (\ref{92}) are obtained by the substitution of
$Q_{\mu\nu\alpha\beta}^{\infty}$, Eq.~(\ref{84}), into
Eqs.~(\ref{86}) and (\ref{87}). Performing
straightforward calculations for the interacting and the noninteracting
systems we arrive at the following results for the high frequency
parts of the xc elastic moduli  
\begin{eqnarray}\nonumber
K_{\text{xc}}^{\infty} &=&
\frac{d+2}{d}\Big[\frac{2}{d}E_{\text{kin}}^{{\text{xc}}}\\ 
&+& \sum_{{\bf k}}\frac{k^{2}\bar{w}''_{k} 
+ (3d+1)k\bar{w}'_{k} + 2d^{2}\bar{w}_{k}}{2d(d+2)}
 G_{2}^{\text{eq}}(k) \Big],\qquad
\label{93}
\end{eqnarray}
\begin{equation}
\mu_{\text{xc}}^{\infty} =
\frac{2}{d}E_{\text{kin}}^{{\text{xc}}}  
+ \sum_{{\bf k}}
\frac{k^{2}\bar{w}''_{k} + (d+1)k\bar{w}'_{k}}{2d(d+2)}
G_{2}^{\text{eq}}(k),
\label{94}
\end{equation}
where
$E_{\text{kin}}^{{\text{xc}}}=E_{\text{kin}}-E_{\text{kin}}^{(0)}$ is
the xc kinetic energy of the equilibrium system. In the special case
of Coulomb interaction the momentum
integrals in Eqs.~(\ref{93}), (\ref{94}) can be expressed in terms of
the potential energy per unit volume, 
$$
E_{\text{pot}}=\frac{1}{2}\sum_{\bf k}\bar{w}_{k}G_{2}^{\text{eq}}(k).
$$  
In $d$ dimensions the Coulomb potential is proportional to
$1/k^{d-1}$. Therefore we get the following identities for the
derivatives, which enter Eqs.~(\ref{93}), (\ref{94})
$$
k\bar{w}'_{k}=-(d-1)\bar{w}_{k} \quad \text{and} \quad 
k^{2}\bar{w}''_{k}=d(d-1)\bar{w}_{k}.
$$
These identities allow us to simplify Eqs.~(\ref{93}), (\ref{94}) as follows 
\begin{eqnarray}
K_{\text{xc}}^{\infty} &=&
\frac{2(d+2)}{d^{2}}E_{\text{kin}}^{{\text{xc}}}
+ \frac{d+1}{d^{2}}E_{\text{pot}},
\label{95}\\
\mu_{\text{xc}}^{\infty} &=&
\frac{2}{d}E_{\text{kin}}^{{\text{xc}}}
- \frac{d-1}{d(d+2)}E_{\text{pot}}.
\label{96}
\end{eqnarray}
The high frequency forms of Eqs.~(\ref{95}), (\ref{96}) are well
known in the literature
\cite{ConVig1999,NifConTos1998,QiaVig2002b}. Commonly they are derived
using the ``third moment sum rule''. Within our formalism the
expressions of Eqs.~(\ref{93}) and (\ref{94}) for
$K_{\text{xc}}^{\infty}$ and $\mu_{\text{xc}}^{\infty}$ come about almost
trivially from the explicit local in time dependence of the stress tensor,
Eq.~(\ref{53}), on the deformation tensor. 

To represent the frequency dependent parts of viscoelastic moduli in
the most convenient
form we decompose the stress tensor operator $\widehat{P}_{\mu\nu}$,
Eq.~(\ref{76}), into a scalar and a traceless parts 
\begin{equation}
\widehat{P}_{\mu\nu} = \delta_{\mu\nu}\widehat{P} + \widehat{\pi}_{\mu\nu},
\label{97}
\end{equation}
where $\widehat{P} = \frac{1}{d}\text{Tr}\widehat{P}_{\mu\nu}$ is the
pressure operator
\begin{equation}
\widehat{P} = \frac{1}{d}\sum_{\bf k}\Big[\frac{k^{2}}{m}
\widetilde{a}^{\dag}_{\bf k}\widetilde{a}_{\bf k}
+ \frac{1}{2}(k\bar{w}'_{k} + d\bar{w}_{k})
\widehat{\widetilde{n}}_{\bf k}\widehat{\widetilde{n}}_{\bf -k}\Big]
\label{98}
\end{equation} 
and $\widehat{\pi}_{\mu\nu}$ is the operator of the traceless part of
the stress tensor ($\text{Tr}\widehat{\pi}_{\mu\nu}=0$)
\begin{equation}
\widehat{\pi}_{\mu\nu} = \sum_{\bf k}
\Big(\frac{k_{\mu}k_{\nu}}{k^{2}} - \frac{\delta_{\mu\nu}}{d}\Big)
\Big(\frac{k^{2}}{m}
\widetilde{a}^{\dag}_{\bf k}\widetilde{a}_{\bf k}
+ \frac{k\bar{w}'_{k} + d\bar{w}_{k}}{2d}
\widehat{\widetilde{n}}_{\bf k}\widehat{\widetilde{n}}_{\bf -k}\Big)
\label{99}
\end{equation} 
Substituting  
$\Delta\widetilde{Q}_{\mu\nu\alpha\beta}(\omega)$ of Eq.~(\ref{81})
into Eqs.~(\ref{86}), (\ref{87}), and using
Eqs.~(\ref{97})--(\ref{99}) we find that 
$\Delta K_{\text{xc}}(\omega)$ equals to the pressure autocorrelation 
function
\begin{equation}
\Delta K_{\text{xc}}(\omega) =
- 2i \int\limits_{0}^{\infty} e^{i\omega t}\left\langle
[\widehat{P}(t),\widehat{P}(0)]
\right\rangle dt,
\label{100}
\end{equation}
whereas the dynamic part of the shear modulus, 
$\Delta \mu_{\text{xc}}(\omega)$, is proportional to the
autocorrelation function of the traceless tensor operator 
$\widehat{\pi}_{\mu\nu}$
\begin{equation}
\Delta \mu_{\text{xc}}(\omega) =
\frac{-2i}{d^{2} + d - 2} 
\int\limits_{0}^{\infty} e^{i\omega t}\left\langle
[\widehat{\pi}_{\mu\nu}(t),\widehat{\pi}_{\mu\nu}(0)]
\right\rangle dt
\label{101}
\end{equation}
The coefficient, $2/(d^{2} + d - 2)$, in
Eq.~(\ref{101}) is exactly the inverse number of independent
components of a second-rank traceless tensor. We would like to outline
a very natural form of Eqs.~(\ref{100}) and (\ref{101}), which is in
clear agreement with the physical significance of the quantities $K$
and $\mu$. 

The frequency dependent contributions to the visco-elastic moduli are
related to the dynamics of the Wigner function and the pair
correlation function (see Eq.~(\ref{80})). In Sec.~VB we have shown
that in the exchange approximation the time-dependent deformation
tensor in the Hamiltonian of Eq.~(\ref{52}) does not induce any
dynamics of $\widetilde{f}({\bf k})$ and $\widetilde{G}_{2}({\bf
  k})$. Therefore only correlations are responsible
for nonvanishing $\Delta K_{\text{xc}}(\omega)$ and 
$\Delta \mu_{\text{xc}}(\omega)$. The most important effect of the
dynamic correlations, which is described by Eqs.~(\ref{100}) and
(\ref{101}), is the memory loss due to collisions. Since in a
zero-temperature Fermi system the collisional dissipation
should be suppressed, there is a hope that the frequency dependent parts of
$K_{\text{xc}}$ and $\mu_{\text{xc}}$
do not substantially influence the dynamics. [We note that this is
not in general true for steady state transport situations, where the
dissipation plays an essential role.] Neglecting 
$\Delta K_{\text{xc}}(\omega)$ and $\Delta \mu_{\text{xc}}(\omega)$ we
get a purely elastic xc stress tensor with the bulk and the shear
moduli defined by Eqs.~(\ref{93}) and (\ref{94}).  Another argument due
to Conti and Vignale \cite{ConVig1999} also shows that for an electron
gas the elastic approximation should work reasonably well. Indeed, the
dissipation effects are absent in the x-only approximation that is
valid in the weak coupling (high density) regime. In the strong
coupling (low density) limit electrons tend to form a Wigner crystal -- the
state where the collisional dissipation also vanishes. Therefore one
naturally expects that at all intermediate densities the purely
elastic approximation should provide a reasonable description of
the dynamic stress.       

\subsection{Nonlinear elastic TDLDA}

The linear VK approximation with purely elastic bulk modulus
$K_{\text{xc}}^{\infty}$, Eq.~(\ref{93}),  and shear modulus
$\mu_{\text{xc}}^{\infty}$,  Eq.~(\ref{94}), allows for a simple
nonlinear extension. In this subsection we derive this nonlinear
elastic TDLDA, and formulate a complete set of self-consistent KS
equations in a convenient for practical applications form.  

\subsubsection{Exchange-correlation stress tensor in the elastic TDLDA}

Elastic TDLDA is based on the assumption that both the Wigner
function and the pair correlation function in the Lagrangian frame
preserve their initial form. This corresponds to the dynamics with
extremely pronounced memory that is not destroyed by
the effects of collisional relaxation. To get the stress tensor for
the system evolving from 
the equilibrium state we have to substitute 
$\widetilde{f}({\bf k},t)=f^{\text{eq}}(n_{0};k)$ and
$\widetilde{G}_{2}({\bf k},t)=G_{2}^{\text{eq}}(n_{0};k)$ into
Eq.(\ref{53}). As a result the stress tensor in the Lagrangian frame
takes the form   
\begin{eqnarray}\nonumber
\widetilde{P}_{\mu\nu} &=& \frac{\delta_{\mu\nu}}{\sqrt{g}}
\frac{2}{d}E_{\text{kin}}(n_{0})
+ \frac{1}{2g}\sum_{\bf k}
\Big[\frac{k_{\mu}k_{\nu}}{\|{\bf k}\|}\bar{w}'(\|{\bf k}\|)\\
&+& g_{\mu\nu}\bar{w}(\|{\bf k}\|)\Big]
G_{2}^{\text{eq}}(n_{0};k).
\label{102}
\end{eqnarray}
Tensor $\widetilde{P}_{\mu\nu}(\bm\xi,t)$ of Eq.~(\ref{102}) locally
depends on the density, $n_{0}(\bm\xi)$, and Green's
deformation tensor, $g_{\mu\nu}(\bm\xi,t)$, in a given point $\bm\xi$
of the Lagrangian space. Transforming this tensor back to the
laboratory frame and subtracting the KS stress tensor of
Eq.~(\ref{66}) we obtain the following result for the xc stress tensor
in the physical ${\bf x}$-space
\begin{eqnarray}\nonumber
P_{\mu\nu}^{\text{xc}} 
&=& \frac{2\bar{g}_{\mu\nu}}{d}\sqrt{\bar{g}}
E_{\text{kin}}^{\text{xc}}\left(\frac{n}{\sqrt{\bar{g}}}\right)
+ \frac{\sqrt{\bar{g}}}{2}\sum_{\bf p}
\Big[\frac{p_{\mu}p_{\nu}}{p}\bar{w}'_{p}\\
&+& \delta_{\mu\nu}\bar{w}_{p}\Big]
G_{2}^{\text{eq}}\left(\frac{n}{\sqrt{\bar{g}}};
\sqrt{\bar{g}^{\alpha\beta}p_{\alpha}p_{\beta}}\right).
\label{103}
\end{eqnarray}
Equation (\ref{103}) determines the xc stress tensor as a function of the
time-dependent density, $n({\bf x},t)$, and Cauchy's deformation
tensor $\bar{g}_{\mu\nu}({\bf x},t)$. Let us remind that the memory
related nonlocality of $P_{\mu\nu}^{\text{xc}}$, Eq.~(\ref{103}), is
hidden in the ``local'' representation of the function 
$n_{0}(\bm\xi({\bf x},t))$ (see Eq.~(\ref{65})). The ``elastic'' xc
potential is the solution to the Poisson Eq.~(\ref{20}), where the xc
``charge density'' is defined after Eq.~(\ref{21}) and (\ref{103}). 

In the exchange approximation the stress tensor $P_{\mu\nu}^{\text{xc}}$,
Eq.~(\ref{103}), reduces to the x-only tensor $P_{\mu\nu}^{\text{x}}$
of Eq.~(\ref{73}) that is exact in the weak coupling limit. In the
linear response regime the corrections to the density and to the Cauchy's
deformation tensor are proportional to the displacement vector
\begin{equation}
n = n_{0} - \nabla n_{0}{\bf u}, \quad
\bar{g}_{\mu\nu} = \delta_{\mu\nu} - 
\frac{\partial u_{\mu}}{\partial x^{\nu}}
- \frac{\partial u_{\nu}}{\partial x^{\mu}}.  
\label{104}
\end{equation}
Linearizing the stress tensor of Eq.~(\ref{103}) and using
Eq.~(\ref{104}) we straightforwardly recover VK approximation
\cite{VigKohn1996,VigUllCon1997} with the elastic moduli
$K_{\text{xc}}^{\infty}$, Eq.~(\ref{93}), and
$\mu_{\text{xc}}^{\infty}$, Eq.~(\ref{94}). 

\subsubsection{Self-consistent Kohn-Sham equations}

Let us formulate the complete set of self-consistent KS equations in the
elastic TDLDA. The Kohn-Sham formulation of TDDFT allows to calculate
the density 
$n({\bf x},t)$ and the velocity ${\bf v}({\bf x},t)$ in the interacting
$N$-particle system using the ideal gas formulas 
\begin{eqnarray}
n({\bf x},t) &=& \sum_{j=1}^{N}|\phi_{j}({\bf x},t)|^{2}, 
\label{105}\\
{\bf v}({\bf x},t) &=& \frac{1}{n}\sum_{j=1}^{N}
\frac{i}{2m}\left[\phi_{j}\nabla\phi_{j}^{*} 
- \phi_{j}^{*}\nabla\phi_{j}\right].
\label{106}
\end{eqnarray}
Single particle orbitals $\phi_{j}({\bf x},t)$ satisfy the time-dependent
KS equations 
\begin{equation}
i\frac{\partial\phi_{j}}{\partial t} = -\frac{\nabla^{2}}{2m}\phi_{j} +
\left(U_{\text{ext}} 
+ U_{\text{eff}}[n,\bar{g}_{\mu\nu}]\right)\phi_{j},
\label{107}
\end{equation}
where $U_{\text{ext}}({\bf x},t)$ is the external potential. For the
practically important case of 3D system with Coulomb interaction the
effective potential $U_{\text{eff}}[n,\bar{g}_{\mu\nu}]({\bf x},t)$ is
the solution to the following Poisson equation
\begin{equation}
\nabla^{2}U_{\text{eff}} = 4\pi(e^{2}n 
+ \rho_{\text{xc}}[n,\bar{g}_{\mu\nu}]).
\label{108} 
\end{equation}
The first term in the brackets in Eq.~(\ref{108}) generates the
Hartree potential, $U_{\text{H}}$, while the second term is
responsible for the xc potential. The xc ``charge density'',
$\rho_{\text{xc}}$, is the local functional of $n$ and
$\bar{g}_{\mu\nu}$
\begin{equation} 
\rho_{\text{xc}} = \frac{1}{4\pi}\frac{\partial}{\partial x^{\mu}}
\left[\frac{1}{n}
\frac{\partial}{\partial x^{\nu}} 
P^{\text{xc}}_{\mu\nu}(n,\bar{g}_{\mu\nu})\right],
\label{109}
\end{equation}
where $P^{\text{xc}}_{\mu\nu}(n,\bar{g}_{\mu\nu})$ is the 
{\it function} of $n({\bf x},t)$ and $\bar{g}_{\mu\nu}({\bf x},t)$, 
which is defined by Eq.~(\ref{103}). In Appendix~C we show that for a
Coulomb system Eq.~(\ref{103}) simplifies as follows
\begin{equation}
P^{\text{xc}}_{\mu\nu} = \frac{2}{3}\bar{g}_{\mu\nu}\sqrt{\bar{g}}
E_{\text{kin}}^{\text{xc}}\left(\frac{n}{\sqrt{\bar{g}}}\right)
+ L_{\mu\nu}(\bar{g}_{\alpha\beta})
E_{\text{pot}}\left(\frac{n}{\sqrt{\bar{g}}}\right)
\label{110}
\end{equation}
where $L_{\mu\nu}(\bar{g}_{\alpha\beta})$ is a purely geometric factor
that is explicitly defined in Appendix~C. Therefore the dependence of
$P^{\text{xc}}_{\mu\nu}(n,\bar{g}_{\mu\nu})$ on  $\bar{g}_{\mu\nu}$
and on $n/\sqrt{\bar{g}}$ is completely factorized, which should
significantly simplify practical applications. The kinetic, 
$E_{\text{kin}}^{\text{xc}}(n)$, and the potential,
$E_{\text{pot}}(n)$, energies of the homogeneous electron gas can be
expressed in terms of the xc energy per particle, 
$\epsilon_{\text{xc}}(n)$
(see, for example, Ref.~\onlinecite{ConVig1999}). For $d=3$ we get
$$
E_{\text{kin}}^{\text{xc}}(n) = 3n^{\frac{7}{3}}
\left(\frac{\epsilon_{\text{xc}}}{n^{\frac{1}{3}}}\right)', \quad
E_{\text{pot}}(n) = - 3n^{\frac{8}{3}}
\left(\frac{\epsilon_{\text{xc}}}{n^{\frac{2}{3}}}\right)'
$$
Hence our nonadiabatic TDLDA requires only a knowledge of the function
$\epsilon_{\text{xc}}(n)$ for the homogeneous electron gas, exactly as
the common static LDA does.

The density $n$, which enters Eqs.~(\ref{108})--(\ref{110}), is
related to KS orbitals via Eq.~(\ref{105}). The second basic variable,
Cauchy's deformation tensor $\bar{g}_{\mu\nu}$, is uniquely
determined by the velocity ${\bf v}({\bf x},t)$, Eq.~(\ref{106}). To
compute the deformation tensor we need to solve the trajectory
equation of Eq.~(\ref{24}) and then substitute the solution into the
definition of $\bar{g}_{\mu\nu}$, Eq.~(\ref{62}). It is,
however, more convenient to determine this tensor directly from the
solution of an equation of motion for $\bar{g}_{\mu\nu}({\bf x},t)$
\cite{note3}. This equation of motion can be derived 
as follows. Let us  consider the contravariant tensor
$\bar{g}^{\mu\nu}$ (the inverse of $\bar{g}_{\mu\nu}$)
\begin{equation}
\bar{g}^{\mu\nu}= 
\frac{\partial x^{\mu}}{\partial \xi^{\alpha}}
\frac{\partial x^{\nu}}{\partial \xi^{\alpha}}.
\label{111}
\end{equation}
Using the trajectory equation of Eq.~(\ref{24}) we can compute the
time derivative of $\bar{g}^{\mu\nu}$, Eq.~(\ref{111}), at constant $\bm\xi$
(i.e. within the Lagrangian description)  
\begin{equation}
\left(\frac{\partial\bar{g}^{\mu\nu}}{\partial t}\right)_{\bm\xi} =
\frac{\partial v^{\mu}}{\partial x^{\alpha}}\bar{g}^{\alpha\nu} 
+\bar{g}^{\mu\alpha}\frac{\partial v^{\nu}}{\partial x^{\alpha}}
\label{112}
\end{equation}
The time derivative of $\bar{g}_{\mu\nu}$ can be related to the time
derivative of $\bar{g}^{\mu\nu} = (\bar{g}_{\mu\nu})^{-1}$ as follows
$\partial_{t}\bar{g} = -\bar{g}(\partial_{t}\bar{g}^{-1})\bar{g}$. 
Using this relation and taking into account that 
$$
(\partial_{t})_{\bm\xi} = (\partial_{t})_{\bf x} + {\bf v}\nabla, 
$$
we get the final equation of motion for Cauchy's deformation tensor
$\bar{g}_{\mu\nu}({\bf x},t)$
\begin{equation}
\frac{\partial\bar{g}_{\mu\nu}}{\partial t} =
- v^{\alpha}\frac{\partial\bar{g}_{\mu\nu}}{\partial x^{\alpha}}
- \frac{\partial v^{\alpha}}{\partial x^{\mu}}\bar{g}_{\alpha\nu} 
- \frac{\partial v^{\alpha}}{\partial x^{\nu}}\bar{g}_{\alpha\mu}
\label{113}
\end{equation}
Equation (\ref{113}) should be solved with the initial condition 
$\bar{g}_{\mu\nu}({\bf x},0)=\delta_{\mu\nu}$, which follows from the
initial condition for the trajectory equation of Eq.~(\ref{24}).

The system of Eqs.~(\ref{105})--(\ref{110}), (\ref{113}) constitute the
complete set of self-consistent KS equations in the nonlinear elastic
TDLDA. In the equilibrium situation
($\bar{g}_{\mu\nu}=\delta_{\mu\nu}$) this system reduces to the common
static KS equation with the LDA xc potential. In the linear regime
it recovers the results of VK approximation with the elastic moduli of
Eqs.~(\ref{95}), (\ref{96}). The nonadiabatic memory effects are
described by Cauchy's deformation tensor, which satisfies
Eq.~(\ref{113}). It should be noted that from the computational point
of view the solution of this equation do not introduce any addition
difficulties. Formally Eq.~(\ref{113}) has the same structure as the
time-dependent KS Eq.~(\ref{107}). Hence Eqs.~(\ref{107}) and
(\ref{113}) can be solved simultaneously by the same method.  

Very recently VK approximation has been successfully applied to the
description of optical and polarization properties of many different
systems, such as atoms, molecules, semiconductors and polymers
\cite{deBoeij2001,Faassen2002,Faassen2003,UllBur2004,FaaBoe2004}. Since
VK approximation is a linearized version of our theory, we
hope that the general TDLDA also will become a useful tool for studying
nonlinear time-dependent phenomena.

\section{Conclusion}

TDDFT extends powerful ideology of the ground state DFT to the domain
of nonequilibrium phenomena. However, in contrast to the static DFT,
which is currently a common computational tool in many branches of
physics, its time-dependent counterpart still suffer from a number of
unresolved problems. One of those problems is a lack of well founded basic
local approximation that would play a role similar to LDA in the static
DFT. In this paper we have shown that the local approximation in TDDFT can
be regularly derived, but this derivation requires almost complete
reconsideration of the theory. We reformulated TDDFT from the point of
view of a local co-moving observer. The new formulation of the theory
shows that the most natural basic variables in TDDFT are the local
geometric characteristics of the deformations in a quantum many-body
system. 

Throughout this paper we used the analogy of TDDFT to
the classical continuum mechanics. The importance of the hydrodynamic
interpretation, which perfectly fits the very idea of DFT, is one of
the messages of the present work. Using the hydrodynamic formulation
of TDDFT we were able to relate the xc potentials to the local
stress. In particular we proved that the exact xc force must have
a form of a divergence of a second-rank tensor. The well known zero
force and zero torque sum rules are direct consequences of this strong
local requirement. The functional dependence of xc potential on
the basic variables also acquires a clear physical meaning. It
corresponds to the stress-deformation relation, which is very natural
from the point of view of continuum mechanics. If spatial derivatives
of the deformation tensor are small, the stress-deformation relation
becomes local and therefore we get the local approximation for the xc
potential in TDDFT. It is natural to abbreviate this approximation as
TDLDA, which means time-dependent local deformation approximation. 
In the linear response regime the general
stress-deformation relation (TDLDA) reduces to the linear Hook's law
\cite{LandauVI:e}, which exactly coincides with the visco-elastic VK
approximation \cite{VigKohn1996,VigUllCon1997}. The formal
applicability conditions for the general nonlinear TDLDA are the
same as for the linear VK approximation. It should be also noted
that neither of previously proposed nonlinear phenomenological
constructions \cite{DobBunGro1997,VigUllCon1997,KurBae2004} is in
general confirmed by the present regular microscopic consideration.        

In the last section of this paper we introduced the elastic TDLDA. In
this approximation the xc stress tensor is simply a function of the
density and of the Cauchy's deformation tensor. For
a system with Coulomb interaction we presented the xc stress tensor and
the xc potential in an explicit ``ready for implementation'' form. We
also formulated the full set of self-consistent KS equations in
TDLDA. In the equilibrium state the deformation tensor is diagonal and
TDLDA reduces to the standard static LDA, while in the linear response
regime it recovers VK approximation. To conclude we mention that the
self-consistent equations of Sec.~VD2 can be straightforwardly
reformulated in terms of xc vector potential. The only difference is
that the Poisson equation for $U_{\text{xc}}$ should be replaced by
Eq.~(\ref{18a}) which relates ${\bf A}^{\text{xc}}$ to the xc stress
tensor. This replacement introduces one more evolution equation which
should be solved simultaneously with the KS equation and the equation
for Cauchy's deformation tensor.

\acknowledgments
This work was supported by the Deutsche Forschungsgemeinschaft under
Grant No. PA 516/2-3.

\appendix
\section{Stress tensors in a general noninertial frame}

The microscopic representation for the stress tensor 
$\widetilde{P}_{\mu\nu}(\bm\xi,t)$ in a general local noninertial
frame has been derived in I: 
\begin{equation}
\widetilde{P}_{\mu\nu}(\bm\xi,t) = \widetilde{T}_{\mu\nu}(\bm\xi,t)
+\widetilde{W}_{\mu\nu}(\bm\xi,t),
\label{A1}
\end{equation}
where the kinetic stress tensor, $\widetilde{T}_{\mu\nu}(\bm\xi,t)$, and
the interaction stress tensor, $\widetilde{W}_{\mu\nu}(\bm\xi,t)$,
are obtained by the transformation of $T_{\mu\nu}({\bf x},t)$,
Eq.~(\ref{13}), and $W_{\mu\nu}({\bf x},t)$, Eq.~(\ref{14}), to the new
frame. Namely, 
\begin{eqnarray}
\widetilde{T}_{\mu\nu}(\bm\xi,t) &=&
\frac{\partial x^{\alpha}}{\partial\xi^{\mu}}
\frac{\partial x^{\beta}}{\partial\xi^{\nu}}
T_{\alpha\beta}({\bf x}(\bm\xi,t),t),
\label{A2} \\
\widetilde{W}_{\mu\nu}(\bm\xi,t) &=&
\frac{\partial x^{\alpha}}{\partial\xi^{\mu}}
\frac{\partial x^{\beta}}{\partial\xi^{\nu}}
W_{\alpha\beta}({\bf x}(\bm\xi,t),t).
\label{A3}
\end{eqnarray}
The result of the transformation, Eqs.~(\ref{A2}), (\ref{A3}), takes
the following form
\begin{widetext}
\begin{eqnarray}
\widetilde{T}_{\mu\nu}(\bm\xi,t) &=& \frac{1}{2m}\left\langle
\left(\widehat{K}_{\mu}g^{-\frac{1}{4}}\widetilde{\psi}\right)^{\dag}
\left(\widehat{K}_{\nu}g^{-\frac{1}{4}}\widetilde{\psi}\right) +
\left(\widehat{K}_{\nu}g^{-\frac{1}{4}}\widetilde{\psi}\right)^{\dag}
\left(\widehat{K}_{\mu}g^{-\frac{1}{4}}\widetilde{\psi}\right) -
\frac{g_{\mu\nu}}{2}\frac{1}{\sqrt{g}}
\frac{\partial}{\partial\xi^{\alpha}}\sqrt{g}g^{\alpha\beta}
\frac{\partial}{\partial\xi^{\beta}}
\frac{\widetilde{\psi}^{\dag}\widetilde{\psi}}{\sqrt{g}}\right\rangle,
\label{A4}\\
\widetilde{W}_{\mu\nu}(\bm\xi,t) &=& 
-\frac{g_{\mu\alpha}g_{\nu\beta}}{2\sqrt{g}}
\int_{0}^{1}d\lambda\int d{\bm\eta}d{\bm\eta'}
\delta(\bm\xi - {\bf z}_{\bm\eta,\bm\eta'}(\lambda))
\frac{\dot{z}^{\alpha}_{\bm\eta,\bm\eta'}(\lambda)
\dot{z}^{\beta}_{\bm\eta,\bm\eta'}(\lambda)}{l_{\bm\eta,\bm\eta'}}
\frac{\partial w(l_{\bm\eta,\bm\eta'})}{\partial l_{\bm\eta,\bm\eta'}}
\widetilde{G}_{2}(\bm\eta,\bm\eta').
\label{A5}
\end{eqnarray}
\end{widetext}
Here the function ${\bf z}_{\bm\eta,\bm\eta'}(\lambda)$ is the
geodesic that connects points $\bm\eta$ and $\bm\eta'$, and 
$l_{\bm\eta,\bm\eta'}$ is the length of this geodesic. 
The curve ${\bf z}_{\bm\eta,\bm\eta'}(\lambda)$ can be found from the
solution of the geodesic equation (see, for example,
Ref.~\onlinecite{DubrovinI}) 
\begin{equation}
\ddot{z}^{\mu}(\lambda) + \Gamma^{\mu}_{\alpha\beta}({\bf z})
\dot{z}^{\alpha}(\lambda)\dot{z}^{\beta}(\lambda)=0,
\label{A6}
\end{equation}
supplemented by the boundary conditions 
${\bf z}(0)={\bm\eta}$, ${\bf z}(1)={\bm\eta}'$. In Eqs.~(\ref{A5}) and
(\ref{A6}) $\dot{\bf z}=\partial{\bf z}/\partial\lambda$, and
$\Gamma^{\mu}_{\alpha\beta}$ is affine connection:
\begin{equation}
\Gamma^{\mu}_{\alpha\beta}(\bm\xi) = \frac{1}{2}g^{\mu\nu}\left(
\frac{\partial g_{\nu\alpha}}{\partial \xi^{\beta}} +
\frac{\partial g_{\nu\beta}}{\partial \xi^{\alpha}} -
\frac{\partial g_{\alpha\beta}}{\partial \xi^{\nu}}
\right).
\label{A7}
\end{equation}

Equations (\ref{A4}) and (\ref{A5}) define tensors
$\widetilde{T}_{\mu\nu}$ and $\widetilde{W}_{\mu\nu}$ as
functionals of the microscopic state of the system. Tensor
$\widetilde{T}_{\mu\nu}$ is a linear functional of the one particle
density matrix, 
$\widetilde{\rho}_{1}(\bm\xi,\bm\xi') = 
\langle\widetilde{\psi}^{\dag}(\bm\xi)\widetilde{\psi}(\bm\xi')\rangle$.
Similarly, $\widetilde{W}_{\mu\nu}$ is a linear functional of the pair
correlation function $\widetilde{G}_{2}(\bm\xi,\bm\xi')=
\langle\widetilde{\psi}^{\dag}(\bm\xi)\widehat{\widetilde{n}}(\bm\xi')
\widetilde{\psi}(\bm\xi)\rangle 
- \widetilde{n}(\bm\xi)\widetilde{n}(\bm\xi')$. Therefore
\begin{equation}
\widetilde{P}_{\mu\nu} = \widetilde{T}_{\mu\nu}
+\widetilde{W}_{\mu\nu} =
\widetilde{P}_{\mu\nu}[\widetilde{\rho}_{1},\widetilde{G}_{2}](\bm\xi,t)
\label{A8}
\end{equation}
Equation (\ref{A8}) is the result, which we need for the discussion of
TDDFT in Sec.~IV.    

\section{Stress tensors for a homogeneous deformation}

For a homogeneous
system with $g_{\mu\nu}(\bm\xi,t)=g_{\mu\nu}(t)$ and $\widetilde{v}_{T\mu}=0$
the general expressions, Eqs.~(\ref{A4}) and (\ref{A5}), for the
stress tensors simplify as follows.
Equation~(\ref{A4}) for the kinetic stress tensor of takes the form
\begin{eqnarray}\nonumber
\widetilde{T}_{\mu\nu} &=& \frac{1}{2m\sqrt{g}}\left\langle
\frac{\partial\widetilde{\psi}^{\dag}}{\partial \xi^{\mu}}
\frac{\partial\widetilde{\psi}}{\partial \xi^{\nu}} +
\frac{\partial\widetilde{\psi}^{\dag}}{\partial \xi^{\nu}}
\frac{\partial\widetilde{\psi}}{\partial \xi^{\mu}}\right\rangle\\
&=& - \frac{1}{m\sqrt{g}}
\frac{\partial^{2}\widetilde{\rho}_{1}(\bm\xi)}{
\partial \xi^{\mu}\partial \xi^{\mu}},
\label{B1}
\end{eqnarray}
where $\widetilde{\rho}_{1}(\bm\xi-\bm\xi') = 
\langle\widetilde{\psi}^{\dag}(\bm\xi)\widetilde{\psi}(\bm\xi')\rangle$
is the one particle density matrix for the homogeneous
system. Introducing the Wigner function 
\begin{equation}
\widetilde{f}({\bf k}) = \int e^{-ik_{\mu}\xi^{\mu}}
\widetilde{\rho}_{1}(\bm\xi)d\bm\xi,
\label{B2}
\end{equation} 
we obtain the following final representation for
$\widetilde{T}_{\mu\nu}$
\begin{equation}
\widetilde{T}_{\mu\nu} = \frac{1}{\sqrt{g}}\sum_{\bf k}
\frac{k_{\mu}k_{\nu}}{m}\widetilde{f}({\bf k})
\label{B3}
\end{equation} 

To calculate the interaction stress tensor, Eq.~(\ref{A5}), we need
to solve the geodesic equation of Eq.~(\ref{A7}). For a
homogeneous metrics the solution is a straight line:
\begin{equation}
{\bf z}_{\bm\eta,\bm\eta'}(\lambda) =
\bm\eta + (\bm\eta' - \bm\eta)\lambda.
\label{B4}
\end{equation}  
The length of the geodesic, $l_{\bm\eta,\bm\eta'}$, can be calculated as
follows
\begin{eqnarray}\nonumber
l_{\bm\eta,\bm\eta'} &=& \int_{0}^{1}
\sqrt{g_{\mu\nu}({\bf z})
\dot{z}^{\mu}(\lambda)\dot{z}^{\nu}(\lambda)}d\lambda \\ 
&=& \sqrt{g_{\mu\nu}(\eta^{\mu}-\eta^{\mu})(\eta^{\nu}-\eta^{\nu})}
:= \|\bm\eta - \bm\eta'\|.
\label{B5}
\end{eqnarray}
Substituting Eqs.~(\ref{B4}), (\ref{B5}) into Eq.~(\ref{A5}) and
taking into account that for a homogeneous system 
$\widetilde{G}_{2}(\bm\eta,\bm\eta')=\widetilde{G}_{2}(\bm\eta-\bm\eta')$,
we arrive at the following result
\begin{equation}
\widetilde{W}_{\mu\nu} = 
- \frac{g_{\mu\alpha}g_{\nu\beta}}{2\sqrt{g}}\int
\frac{\xi^{\alpha}\xi^{\beta}}{\|\bm\xi\|}
\frac{\partial w(\|\bm\xi\|)}{\partial\|\bm\xi\|} 
\widetilde{G}_{2}(\bm\xi)d\bm\xi.
\label{B6} 
\end{equation}
Let us expand the pair correlation function,
$\widetilde{G}_{2}(\bm\xi)$, in a Fourier series,
\begin{equation}
\widetilde{G}_{2}(\bm\xi) = \sum_{\bf k}e^{ik_{\mu}\xi^{\mu}}
\widetilde{G}_{2}({\bf k}),
\label{B7}
\end{equation} 
and express $\widetilde{W}_{\mu\nu}$, Eq.~(\ref{B6}), in terms of
$\widetilde{G}_{2}({\bf k})$. 

First we note, that the following simple
relation holds. Let $\bar{F}(|{\bf k}|)$, where 
$|{\bf k}|=\sqrt{k_{\mu}k_{\mu}}$, be the Fourier component of a
function $F(|\bm\xi|)$, i.e.
\begin{equation}
\bar{F}(|{\bf k}|) = \int e^{-ik_{\mu}\xi^{\mu}}F(|\bm\xi|)d\bm\xi.
\label{B8}   
\end{equation}
Then, the Fourier component of the function $F(\|\bm\xi\|)$ can be
expressed in terms of $\bar{F}$ as follows
\begin{equation}
\int e^{-ik_{\mu}\xi^{\mu}}F(\|\bm\xi\|)d\bm\xi =
\frac{1}{\sqrt{g}}\bar{F}(\|{\bf k}\|),
\label{B9}
\end{equation}
where $\|\bm\xi\|=\sqrt{g_{\mu\nu}\xi^{\mu}\xi^{\nu}}$ (see
Eq.~(\ref{B5})), and 
\begin{equation}
\|{\bf k}\| = \sqrt{g^{\mu\nu}k_{\mu}k_{\nu}}
\label{B10}
\end{equation}
Substituting the expansion of
Eq.~(\ref{B7}) into Eq.~(\ref{B6}) and using Eqs.~(\ref{B8}),
(\ref{B9}), we get the required representation for the interaction 
stress tensor
\begin{equation}
\widetilde{W}_{\mu\nu} = \frac{1}{2g}\sum_{\bf k}\left[
\frac{k_{\mu}k_{\nu}}{\|{\bf k}\|}\bar{w}'(\|{\bf k}\|)
+ g_{\mu\nu}\bar{w}(\|{\bf k}\|)\right]\widetilde{G}_{2}({\bf k}).
\label{B11}
\end{equation} 
In Eq.~(\ref{B11}) the function $\bar{w}(|{\bf k}|)$ is the Fourier
component of the interaction potential, $w(|{\bm\xi}|)$, and 
$\bar{w}'(x)=d\bar{w}(x)/dx$. 

The stress tensor $\widetilde{P}_{\mu\nu}$ of Eq.~(\ref{53}) is the sum
of $\widetilde{T}_{\mu\nu}$, Eq.~(\ref{B3}), and
$\widetilde{W}_{\mu\nu}$, Eq.~(\ref{B11}).   

\section{Elastic stress tensor in Coulomb systems}

The general expression for xc stress tensor, Eq.~(\ref{103}), can be
represented in a much more simple form if the particles interact via
Coulomb potential, $\bar{w}_{p}=A_{d}/p^{d-1}$ (where $A_{3}=4\pi
e^{2}$ and $A_{2}=2\pi e^{2}$). Below we show that in this case the
second term in Eq.~(\ref{103}) can be related to the
potential energy, $E_{\text{pot}}$, of a homogeneous electron gas with
the density $n/\sqrt{\bar{g}}$. 

Let us represent the momentum integral in Eq.~(\ref{103}) as a sum of two
terms
\begin{equation}
W_{\mu\nu} = W_{\mu\nu}^{(1)} + W_{\mu\nu}^{(2)}
\label{C1}
\end{equation}
where
\begin{eqnarray}
W_{\mu\nu}^{(1)} &=& \delta_{\mu\nu}
\frac{\sqrt{\bar{g}}}{2}\sum_{\bf p}\frac{A_{d}}{p^{d-1}}
G_{2}^{\text{eq}}\left(
\sqrt{\bar{g}^{\alpha\beta}p_{\alpha}p_{\beta}}\right),
\label{C2}\\
W_{\mu\nu}^{(2)} &=& -(d-1)\frac{\sqrt{\bar{g}}}{2}\sum_{\bf p}
A_{d}\frac{p_{\mu}p_{\nu}}{p^{d+1}}
G_{2}^{\text{eq}}\left(
\sqrt{\bar{g}^{\alpha\beta}p_{\alpha}p_{\beta}}\right).
\label{C3}
\end{eqnarray}
To shorten the notations we retain only important momentum dependence
in the argument of the pair correlation function $G_{2}^{\text{eq}}$. 

Transformation of $W_{\mu\nu}^{(1)}$, Eq.~(\ref{C2}), is
straightforward. By changing the integration variables this equation
can be reduced to the form
\begin{equation}
W_{\mu\nu}^{(1)} = \delta_{\mu\nu}
\frac{\bar{g}}{2}\sum_{\bf p}
\frac{A_{d}}{[\bar{g}_{\alpha\beta}p_{\alpha}p_{\beta}]^{\frac{d-1}{2}}}
G_{2}^{\text{eq}}(p)
\label{C4}
\end{equation}
Separation the integration over the modulus and the direction of
momentum in Eq.~(\ref{C4}) yields the following result for 
$W_{\mu\nu}^{(1)}$
\begin{equation}
W_{\mu\nu}^{(1)} = \delta_{\mu\nu}\bar{g}
\left\langle
\frac{1}{[\bar{g}_{\alpha\beta}l^{\alpha}l^{\beta}]^{\frac{d-1}{2}}}
\right\rangle_{\bf l}E_{\text{pot}},
\label{C5}
\end{equation}
where ${\bf l}$ is a unit vector (${\bf l}^{2}=1$), and the angle
brackets, $\langle(...)\rangle_{\bf l}$, stand for the averaging over
the directions of ${\bf l}$.

The momentum integral for $W_{\mu\nu}^{(2)}$, Eq.~(\ref{C3}), can be
reduced to a similar form. Let us first represent the deformation
tensor $\bar{g}_{\mu\nu}$ in terms of its eigenvalues, $\lambda_{j}^{2}$,
and eigen vectors, $\eta_{j\mu}$,
\begin{equation}
\bar{g}_{\mu\nu} = \lambda_{j}^{2}\eta_{j\mu}\eta_{j\nu}.
\label{C6}
\end{equation}   
Here $j=1,\dots,d$ labels the eigen vectors $\bm\eta_{j}$ that
satisfy the completeness and the orthonormality conditions
\begin{equation}
\eta_{j\mu}\eta_{j\nu} = \delta_{\mu\nu}, \quad
\eta_{i\mu}\eta_{j\mu} = \delta_{ij}.
\label{C7}
\end{equation}   
Tensor $\bar{g}^{\mu\nu}$ has the same eigenvectors, while its
eigenvalues equal to $1/\lambda_{j}^{2}$. Substituting the eigen
vector expansion of $\bar{g}^{\mu\nu}$ into Eq.~(\ref{C3}), and
performing an obvious change of the integration variables we arrive at
the following result for the tensor $W_{\mu\nu}^{(2)}$
\begin{equation}
W_{\mu\nu}^{(2)} = -\bar{g}\eta_{j\mu}\eta_{j\nu}\left\langle
\frac{(d-1)\lambda_{j}^{2}(\bm\eta_{j}{\bf l})^{2}}{
[\bar{g}_{\alpha\beta}l^{\alpha}l^{\beta}]^{\frac{d+1}{2}}}
\right\rangle_{\bf l}E_{\text{pot}},
\label{C8}
\end{equation}
Combining Eq.~(\ref{C1}), (\ref{C5}) and (\ref{C8}) we obtain the
interaction stress tensor $W_{\mu\nu}$ in the following form
\begin{equation}
W_{\mu\nu} = L_{\mu\nu}(g_{\alpha\beta})E_{\text{pot}},
\label{C9}
\end{equation}
where the calculation of the function $L_{\mu\nu}(g_{\alpha\beta})$
involves only the angle integration. The angle integrals (factors with
angle brackets in Eqs.~(\ref{C5}) and (\ref{C8})) are the scalar
functions which depend only on eigen values of $\bar{g}_{\mu\nu}$. For
$d=2,3$ these integrals are reducible to a combination of the standard
elliptic integrals.


\begin{thebibliography}{47}
\bibitem[{not({\natexlab{b}})}]{note1}
\bibinfo{note}{The existence of a KS system requires that the velocity/density
  in the interacting system is noninteracting v-representable. As usual this is
  assumed to be the case}.
\end{thebibliography}

\end{document}